\documentclass[reprint,
amsmath,amssymb,
aps,natbib
]{revtex4-2}

\usepackage{amsmath}
\usepackage{balance}
\usepackage[colorlinks,linkcolor=blue,anchorcolor=blue,citecolor=red]{hyperref}
\usepackage{longtable}
\usepackage{subfigure}
\usepackage{graphicx}
\usepackage{dcolumn}
\usepackage{bm}%
\newcommand{\bea}{\begin{eqnarray}}\newcommand{\eea}{\end{eqnarray}}

\begin{document}

\title{Circular orbits and thin accretion disk around a quantum corrected black hole}
\author{Yu-Heng Shu}
\author{Jia-Hui Huang}
\email{huangjh@m.scnu.edu.cn}
\affiliation{Key Laboratory of Atomic and Subatomic Structure and Quantum Control (Ministry of Education), Guangdong Basic Research Center of Excellence for Structure and Fundamental Interactions of Matter, School of Physics, South China Normal University, Guangzhou 510006, China}
\affiliation{Guangdong Provincial Key Laboratory of Quantum Engineering and Quantum Materials, Guangdong-Hong Kong Joint Laboratory of Quantum Matter, South China Normal University, Guangzhou 510006, China}

\begin{abstract}
In this paper, we fist consider the shadow radius of a quantum corrected black hole proposed recently, and provide a bound on the correction parameter based on the observational data of Sgr A*. Then, the effects of the correction parameter on the energy, angular momenta and angular velocities of particles on circular orbits in the accretion disk are discussed. It is found that the correction parameter has significant effects on the angular momenta of particles on the circular orbits even in the far region from the black hole.
It would be possible to identify the value of the correction parameter by the observations of the angular momenta of particles in the disk. 
It is also found that the radius of the innermost stable circular orbit increase with the increase of the correction parameter, while the radiative efficiency of the black hole decreases with the increase of the correction parameter. Finally, we consider how the correction parameter affect the emitted and observed radiation fluxes from a thin accretion disk around the black hole. Polynomial fitting functions are identified for the relations between the maxima of three typical radiation fluxes and the correction parameter.   		
\end{abstract}
	

\maketitle
	
	
\section{Introduction}

Black holes are mysterious objects predicted by Einstein's general relativity. A typical feature of a black hole is a spacetime singularity surrounded by an event horizon.  
It was also proved that spacetime singularities can not be avoided under certain common physical conditions in general relativity \cite{Penrose:1964wq,Hawking:1970zqf}.
However, singularities may lead to some deep theoretical puzzles, such as black hole information paradox \cite{Marolf:2017jkr,Buoninfante:2024oxl}. Thus, efforts are made to cure  spacetime singularities of black holes, and typical approaches include string theory, loop quantum gravity, etc. (see \cite{Buoninfante:2024oxl} and references therein) .

A particular approach to modifying Einstein's general relativity is the Hamiltonian constraints formulation\cite{Thiemann:2007pyv,Ashtekar:2004eh,Addazi2022}. A longstanding issue in this formulation is how to keep the general diffeomorphism covariance. 
Recently, the issue has been addressed for spherically symmetric vacuum gravity, and the concept of general covariance is formulated into precise equations, resulting in two quantum modified spherical black holes \cite{Zhang:2024khj}. 
Another quantum corrected spherical black hole metric was proposed in loop quantum gravity by studying the gravitational collapse of spherically symmetric dust matter \cite{Lewandowski:2022zce,Kelly:2020lec,Yang:2022btw}. 

The quasinormal modes of these quantum corrected spherical black holes have been studied by several recent works \cite{Yang:2022btw,Cao:2024oud,Zhang:2024svj,Skvortsova:2024atk,Zinhailo:2024kbq,Gong:2023ghh,Malik:2024nhy,Konoplya:2024lch}. The test of the correspondence between grey-body
factors and quasinormal modes of these black holes were studied in \cite{Skvortsova:2024msa}. Meanwhile, the characteristics of shadows and photon rings of these quantum black holes were studied in \cite{Zhang:2023okw,Ye:2023qks,Liu:2024soc,Peng:2020wun}, and the constraints on the quantum correction parameter from Event Horizon Telescope (EHT) observation  \cite{Konoplya:2024lch,Zhao:2024elr} and gravitational wave observation \cite{Zi:2024jla} were also discussed. Furthermore, the image of the accretion disk and the strong gravitational lensing effects of the quantum black holes were explored in \cite{You:2024jeu, Liu:2024wal,Li:2024afr}. Although these extensive studies have explored different characteristics of the recently proposed quantum corrected black holes, it will be interesting to investigate other aspect of the quantum black holes.

The accretion disk around a black hole is commonly believed to be composed of gases swirling toward the black hole from some nearby companion stars. The most popular
accretion disk models are geometrically thin disk model \cite{Shakura:1972te,Page:1974he,Thorne:1974ve}. These thin disk models are relatively simple, especially, the mass accretion rate is supposed to be constant in time and does not depend on the radius of the disk, and most of the gravitational energy is released by radiation which generates the luminosity of the disk.   
The electromagnetic spectrum of the accretion disk around a black hole could be used to determine the characteristics of the central black hole and the underlying gravity theories \cite{Kong:2014wha,Pun:2008ua,Harko:2009rp,Harko:2009kj,Harko:2010ua,Chakraborty:2014eha,Pun:2008ae,Li:2004aq}.

In this paper, we investigate the quantum corrected black hole proposed in \cite{Zhang:2024khj} from several aspects. In Sec.II,  we consider the shadow radius of the black hole and obtain a constraint on the quantum correction parameter. In Sec.III, we study the effects of the correction parameter on the properties of particles on circular orbits around the black hole and the radiative efficiency of the black hole. In Sec.IV, we consider a thin disk model around the black hole and discuss the impact of the correction parameter on the emitted and observed radiation energy fluxes. The final section is devoted to the summary.

\section{Black hole shadow and constraints from observational data}
The quantum corrected black hole model we are interested in this work is described by \cite{Zhang:2024khj}
\begin{equation} \label{metric}  
		ds^{2}=-f(r)dt^{2}+\frac{1}{f(r)}dr^{2}+r^{2}(d\theta^{2}+\sin^{2}\theta d\varphi^{2}),
\end{equation}
where 
\bea
f(r)=(1-\frac{2M}{r})\left(1+\frac{\xi^{2}}{r^{2}}(1-\frac{2M}{r})\right),
\eea 
$M$ is the mass of the black hole and $\xi$ is the quantum correction parameter. Without loss of generality, we take $M=1$ in the following discussion. 
	
For the calculation of the shadow radius of a spherical black hole, many methods have been proposed and for a recent review, we refer \cite{Perlick:2021aok}. 
For a static spherically symmetric metric, the radius $r_{\text{ph}}$ of the photon ring around the black hole is determined by the following equation \cite{Perlick:2021aok}
	\begin{equation}\label{}
		r \omega^{\prime}(r)=\omega(r),
	\end{equation}
	where $\omega(r)=\sqrt{f(r)}$ .
The radius of the black hole shadow observed by an observer at infinity is \cite{EHT:2020qrl}
	\begin{equation}\label{}
		R_{\text{sh}}=\frac{r_{\text{ph}}}{\sqrt{f(r_{\text{ph}})}}.
	\end{equation}
Plugging the metric \eqref{metric} into the above equations, we obtain that the radius of the photon ring is $r_{\text{ph}}=3$ and the shadow radius of the quantum corrected black hole is	
	\begin{equation}\label{}
		R_{\text{sh}}=\frac{27}{\sqrt{27+\xi^{2}}}
	\end{equation} 
	
The EHT captured an image of the supermassive black hole situated at the center of the M87* galaxy in 2019 \cite{EHT-M87-1,EHT-M87-4,EHT-M87-5,EHT-M87-6}. 
 The EHT results reveal that the angular diameter of the shadow of M87* is $\theta_{\text{sh}}=(42\pm3)\mu as$, the distance from M87* to the Earth is $D=(16.8\pm 0.8)$ \text{Mpc} and the estimated mass of M87* is $(6.5\pm0.7)\times10^{9} M_{\odot}$. In 2022, the EHT collaboration reported the shadow results of Sgr A* located in the center of the Milky Way \cite{EHT-SgrA-1,EHT-SgrA-3,EHT-SgrA-4,EHT-SgrA-5,EHT-SgrA-6}, which reveal that the angular diameter of the shadow of Sgr A* is $\theta_{\text{sh}}=(48.7\pm7.0)\mu as$, the distance from Sgr A* to the Earth is $D=8$ \text{kpc} and the estimated mass of Sgr A* is $(4.0^{+1.1}_{-0.6})\times10^{6} M_{\odot}$. These data may lead to constraints on different black hole models \cite{Afrin2023,Vagnozzi2023,EHT2020qrl,Khodadi:2024ubi}. 

The constraint on quantum correction parameter $\xi$ in Eq.\eqref{metric} from data of Sgr A*  was considered in \cite{Konoplya:2024lch}, which is 
\bea
0\leq\xi\lesssim 2.866~ (1\sigma).
\eea
Here we consider the constraint on $\xi$ based on the data of M87*. The diameter of the shadow normalized by the gravitational radius of  M87* is
\bea
d_{\text{M87}}=\frac{\theta_{\text{sh}} D_{\text{M87}}}{M_{\text{M87}}}\lesssim11.0\pm1.5.
\eea 
Then, the shadow radius $R_{\text{sh}}$ should satisfy the following relation within $1\sigma$	
	\begin{equation}\label{}
		4.75\lesssim R_{\text{sh}}\lesssim 6.25,
	\end{equation}
which leads to the following constraint on the quantum correction parameter,
		\begin{equation}\label{xibound}
		0\leq\xi\lesssim2.304~(1\sigma).
	\end{equation}
It is obvious that different observational data will impose different constraints on the radius of the black hole shadow. It seems that the observational data of M87*
lead to a little stronger constraint on the parameter $\xi$. In this work, we always adopt this constraint. 
		
\section{Analysis of particle motion in the equatorial plane}
In this section we focus on the particle motion in the equatorial plane, so we fix the coordinate $\theta=\frac{\pi}{2}$. Then the metric on the equatorial plane can be obtained from Eq.\eqref{metric}, which is
\begin{equation}\label{}
		ds^{2}=-f(r)dt^{2}+\frac{1}{f(r)}dr^{2}+r^{2}d\varphi^{2}. 
\end{equation}
The Lagrangian of a test particle moving in the equatorial plane takes the following form
	\begin{equation}\label{}
		\mathcal L=\frac{1}{2}m\left(-f(r)\dot{t}^{2}+\frac{\dot{r}^{2}}{f(r)}+r^{2}\dot{\varphi}^{2}\right),
	\end{equation}
where $m$ is the mass of the particle and the dot means derivative with respect to the affine parameter for the particle. Then the canonical momenta of the particle can be obtained,	
	\begin{align}
		P_{t}&=\frac{\partial \mathcal L}{\partial \dot{t}}=-mf(r)\dot{t},\\
		P_{r}&=\frac{\partial \mathcal L}{\partial\dot{r}}=\frac{m}{f(r)}\dot{r},\\
		P_{\varphi}&=mr^{2}\dot{\varphi}.
	\end{align}
With the two obvious Killing vectors of the black hole metric, $\partial_t$ and $\partial_\varphi$, the conserved energy $E$ and angular momentum $L$ of the particle is 
\bea\label{E}
-(\partial_t)^t P_t&=&E=mf(r)\dot{t},\\ \label{L}
(\partial_\varphi)^\varphi P_{\varphi}&=&L=mr^{2}\dot{\varphi}.
\eea 
The Hamiltonian of the moving particle is
\bea
\mathcal H= \dot{t}P_t+\dot{r}P_r+\dot{\varphi}P_\varphi-\mathcal{L}=\mathcal L.
\eea
Take into account the conserved energy and angular momentum in Eqs.\eqref{E}\eqref{L}, and the normalization condition of the four-velocity of the particle,  the Hamiltonian of the particle can be written as,
	\begin{align}\label{hamiltonian}
		\mathcal H&=\frac{1}{2}m\left(-\frac{E^{2}}{m^2f(r)}+\frac{1}{f(r)}\dot{r}^{2}+\frac{L^{2}}{m^2r^{2}}\right)=-\frac{m}{2}.
	\end{align}
From the above equation, we can see that the radial motion of the test particle is determined by the energy per mass and angular momentum per mass, so without loss of generality, we take $m=1$ from now on. Then, we have 
	\begin{align}
		&\frac{1}{2}\dot{r}^{2}+\frac{1}{2}f(r)(\frac{L^{2}}{r^{2}}+1)=\frac{1}{2}E^{2} \label{14}
	\end{align}
The effective potential for the particle moving in the radial direction can be defined as
\bea
V_{\text{eff}}=\frac{1}{2}f(r)(\frac{L^{2}}{r^{2}}+1).
\eea

The radius of a circular orbit of the test particle needs to satisfy the following conditions 
\bea\label{circular}
V_{\text{eff}}(r)=\frac{1}{2}E^2,~~\partial_{r}V_{\text{eff}}(r)=0.
\eea
Solving the above two equations, we can obtain the explicit relations between $E,L$ and radius $r$ of a circular orbit,
\bea
E=\sqrt{\frac{2f^2(r)}{2 f(r)-r f'(r)}},~L=\sqrt{\frac{r^3 f'(r)}{2 f(r)-r f'(r)}}.
\eea
For later use, we also calculate the angular velocity of the particle on an orbit with radius $r$, which is
\bea
\Omega=\sqrt{\frac{f'(r)}{2r}}.
\eea
Commonly, not all circular orbits are stable. An interesting critical circular orbit is the innermost stable circular orbit (ISCO), which is determined by the equations in \eqref{circular} and an additional equation $\partial_{r}^{2}V_{\text{eff}}=0$. Thus, the radius $R_{I}$ of the ISCO satisfies the following equation
\bea\label{iscoeq}
\frac{2 r f(r) f''(r)-4 r f'(r)^2+6 f(r) f'(r)}{2 r f(r)-r^2 f'(r)}=0.
\eea

         \begin{figure*}[htbp]
		\centering
		\subfigure[]{
		 \includegraphics[scale=0.42]{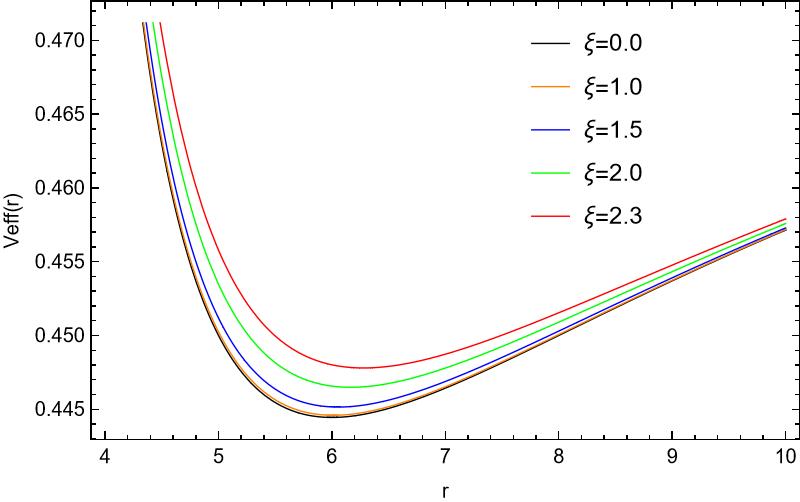}
		  \label{vfa}
		}
		\subfigure[]{
		 \includegraphics[scale=0.42]{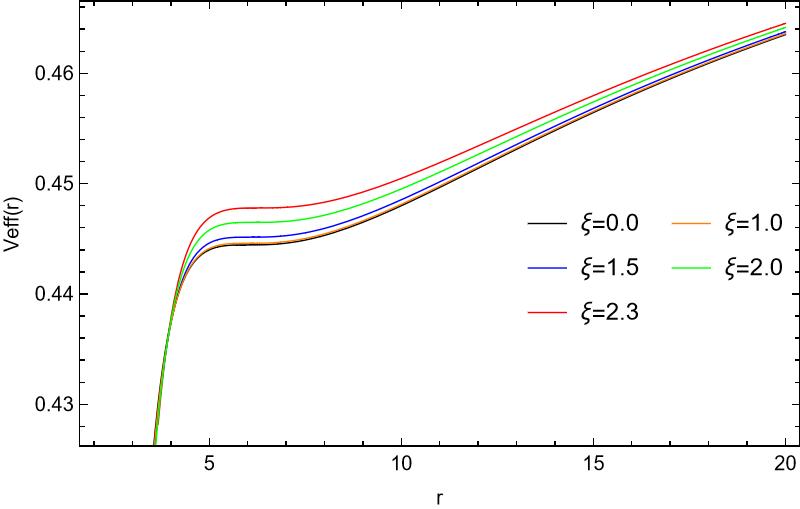}\label{vfb}}

		\caption{(a): The effective potential for a particle on a circular orbit with radius $r$; (b): The Effective potential for a particle moving off the ISCO by certain perturbation. Curves are plotted for different values of $\xi$. }
	    \end{figure*}
	
In Fig.\ref{vfa}, we plot the effective potential as a function of the orbit radii for particles on different circular orbits. One can see that there exists a minimum for each curve with a chosen parameter $\xi$. We only focus on the value of $\xi$ satisfying the bound \eqref{xibound}. The corresponding radius value of the minimum indicates the position of the ISCO. It is easy to check that an orbit is stable when $r>R_{I}$ and is unstable when $r<R_{I}$.  Compared with the Schwarzschild case ($\xi=0$), the effective potential has visible shift only when $\xi>1$. In Fig.\ref{vfb}, we illustrate the effective potential for a particle which is initially located on the ISCO and then leave its orbit for some radial perturbation. The angular momentum of the particle is conserved. If the particle is outgoing from the ISCO, the potential energy increases and the radial motion of the particle slows down. If the particle is ingoing from the ISCO, after a shot steady motion, the effective potential decreases rapidly and the particle rush towards the black hole.

\begin{figure}[h]
		\includegraphics[scale=0.55]{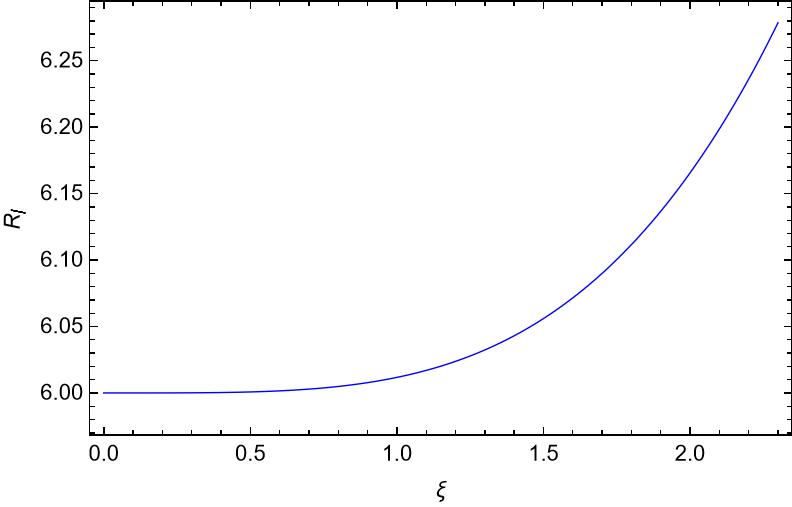}
		\caption{The radius $R_I$ of ISCO of the quantum corrected black hole vs. the correction parameter $\xi$.}
		\label{isco}
	\end{figure}

Plugging the explicit expression of $f(r)$ into Eq.\eqref{iscoeq}, we derive the equation satisfied by $R_{I}$,	
\bea\label{eq:wideeq}
&&r^6(r-6)+3 \xi^2 r^3(r^2-8 r+12)\nonumber\\
&&-2 \xi^4 (r-2)^2 (2 r^2-13 r+24)=0
\eea
This equation can be solved numerically. In Fig. \ref{isco}, we plot $R_{I}$ with respect to the correction parameter $\xi$.
It is obvious that only when $\xi>1$ there is a visible shift for the value of $R_{I}$ from $6$ (Schwarzschild case). 
The monotonically increasing trend of $R_{I}$ with respect to $\xi$ is also clear here. 

	\begin{figure*}[htb]
	\centering
	\subfigure[]{
		\begin{minipage}[t]{0.3\textwidth}
			\centering
			\includegraphics[scale=0.4]{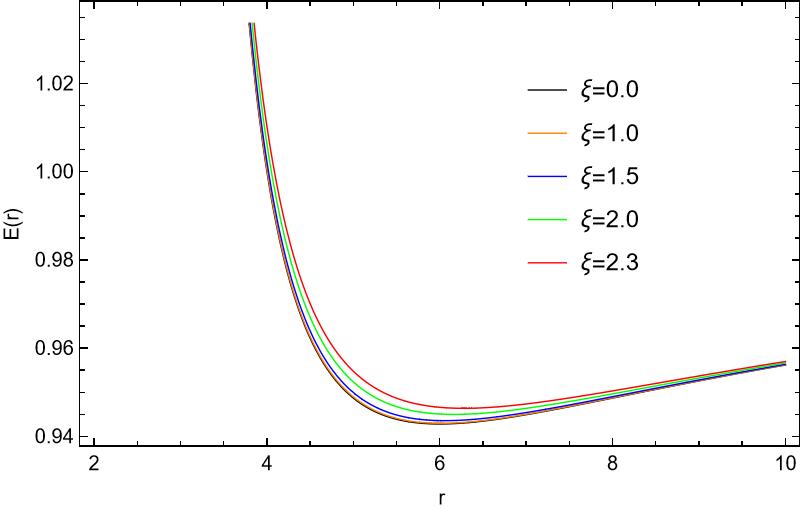}
		\end{minipage}
	}
	\hfill
	\centering
	\subfigure[]{
		\begin{minipage}[t]{0.3\textwidth}
			\centering
			\includegraphics[scale=0.4]{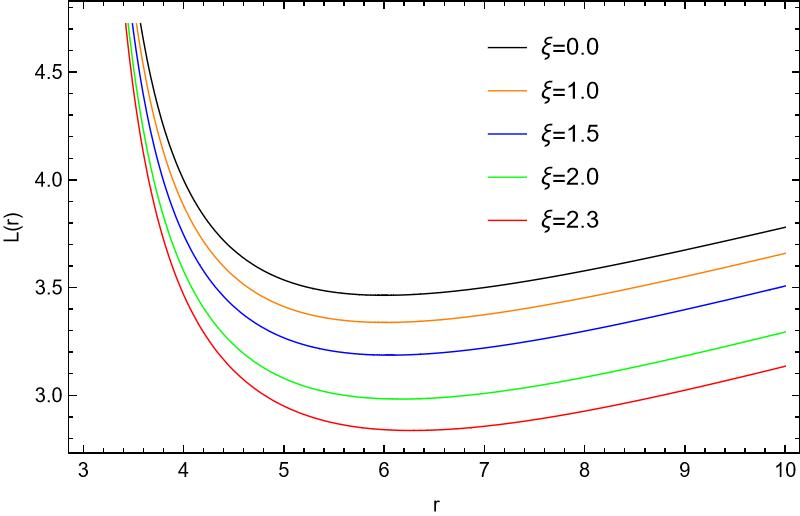}
		\end{minipage}
	}
	\centering
	\hfill
	\subfigure[]{
		\begin{minipage}[t]{0.3\textwidth}
			\centering
			\includegraphics[scale=0.4]{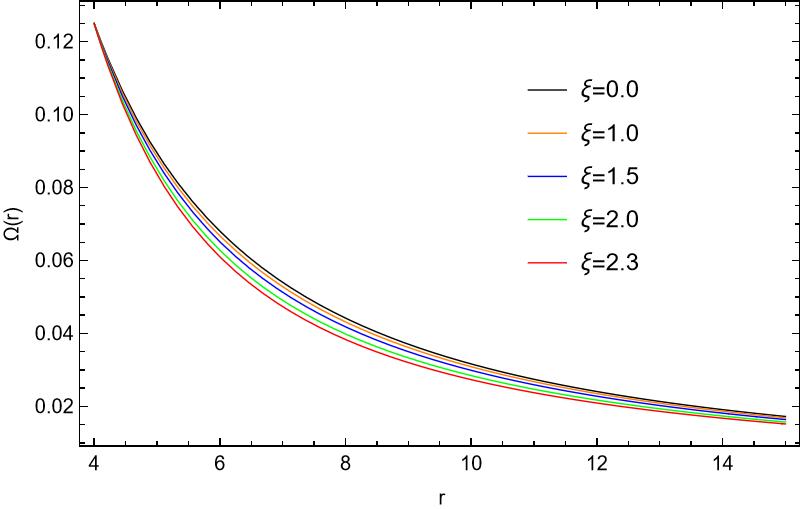}
		\end{minipage}
	}

	\caption{The energy $E$, angular momentum $L$, angular velocity $\Omega$ for a particle on circular orbit with radius $r$ around black holes with different values of $\xi$.}
	\label{e-o-l}
\end{figure*}

	\begin{figure*}[htb]
	\centering
\subfigure[]{
	\begin{minipage}[t]{0.3\textwidth}
		\centering
		\includegraphics[scale=0.42]{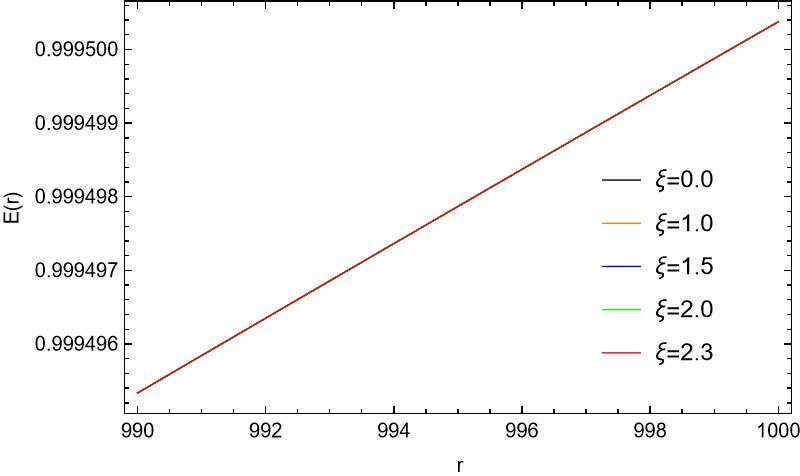}
	\end{minipage}
}
\hfill
	\centering
	\subfigure[]{
		\begin{minipage}[t]{0.3\textwidth}
			\centering
			\includegraphics[scale=0.4]{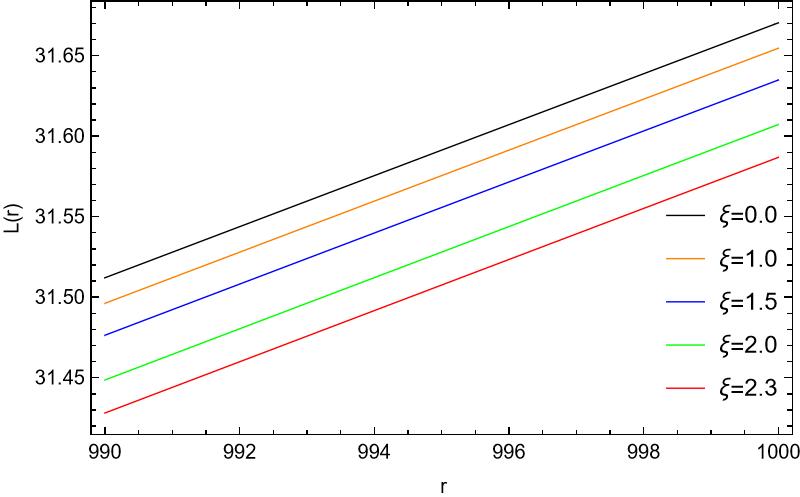}
		\end{minipage}
	}
	\hfill
	\subfigure[]{
		\begin{minipage}[t]{0.3\textwidth}
			\centering
			\includegraphics[scale=0.42]{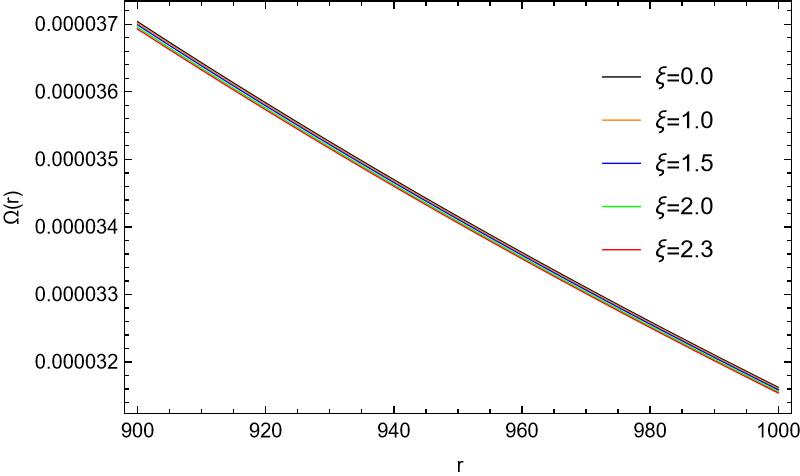}
		\end{minipage}
	}
	\centering
	\caption{ The impact of the parameter $\xi$ on the asymptotic behaviours of the energy $E$, angular momentum $L$ and angular velocity $\Omega$ in far region.}
	\label{e-o-l1}
\end{figure*}

		In Fig.\ref{e-o-l}, the conserved energy $E$, angular momentum $L$, and the angular velocity $\Omega$ for particles on different circular orbits are shown as functions of orbit radius $r$. For particles on different stable circular orbits, the farther away they are from the black hole, the more conserved energy and angular momenta they have. However,  the angular velocity decreases with increasing $r$. These properties lead to the transfer of angular momentum and energy from interior to exterior between particles in the accretion disk. For different values of the correction parameter, we can see that the energy and angular velocity of a particle on an orbit tend to coincide when the orbit is far from the black hole. But, the angular momentum of a particle on an orbit is significantly different for cases with different correction parameters. The reason for these observations is that the asymptotic behaviours of the energy $E$ and angular velocity $\Omega$ have no relation with $\xi$, while angular momentum $L$ has. This is illustrated in Fig.\ref{e-o-l1}.  
So, the correction parameter $\xi$ would be examined in observations by measuring the angular momentum distribution in the accretion disk.

An important quantity related with energy of particles on ISCO is the radiative efficiency of a black hole, which characterizes the ratio between the energy radiated away
during the accretion process and the initial energy of the accretion matter \cite{Page:1974he,Collodel:2021gxu,Wu:2024sng,Kurmanov:2024hpn,Liu:2024brf,Lee:2022rtg,Asukula:2023akj,Olmo:2023lil}. 
Consider that a particle of unit mass goes from infinity to the ISCO, and assume that all lost energy is converted to radiative energy and reaches to infinity.  	 
Then, the radiative efficiency $\epsilon$ is defined as 
	    \begin{align}
		\epsilon=\frac{E_{\infty}-E_{\text{ISCO}}}{E_{\infty}}\approx 1-E_{\text{ISCO}},
		\end{align}
where we use  $E_{\infty}\approx1$. 		
In Tab.\ref{1}, we chose some specific values for the parameter $\xi$, and calculate the radii of the ISCOs, the energy of particles on the ISCOs and the radiative efficiencies.
One can observe that the radiative efficiency is decreasing with the increase of the parameter $\xi$.

	\begin{table}[h]
		\renewcommand{\arraystretch}{1.5}
	\caption{\label{1}
 The radius $R_{I}$ of the ISCO, the energy $E_{\text{ISCO}}$ of a particle on the ISCO and the radiative efficiency $\epsilon$ of the black hole for different values of $\xi$.}
	\begin{ruledtabular}
		\begin{tabular}{cccc}
			\mbox{ $\xi$}&\mbox{$R_{I}$}&\mbox{$E_{\text{ISCO}}$}&\mbox{$\epsilon(\%)$}\\
			\hline
			0&6&0.942809&5.7191\\
			
			0.5&6.00076&0.942819& 5.71809\\
			
			1&6.01171 &0.942964&5.70355 \\
			1.5&6.05598&0.943554&5.64458\\
			2&6.16559&0.944976&5.50244\\
			2.304&6.2803&0.94638&5.36199
		\end{tabular}
	\end{ruledtabular}
    \end{table}

\section{Radiation flux of a thin disk}
In this section we focus on the properties of radiation flux by a thin accretion disk around the quantum corrected black hole. In the spherical coordinates, the radiation flux emitted by the disk is defined as \cite{Page:1974he,Collodel:2021gxu,Wu:2024sng},
	\begin{equation}\label{flux}
		F=-\frac{\dot{M}}{4\pi\sqrt{-g/g_{\theta\theta}}}\frac{\Omega_{,r}}{(E-\Omega L)^{2}}\int_{R_{I}}^{r}(E-\Omega L)L_{,r}dr,
	\end{equation}
	where $\dot{M}$ is the mass accretion rate, $g$ is the determinant of the metric. 
Due to the strong gravitational field around the black hole and the high-speed movement of the particles on the accretion disk, the radiative photons experience 
the effects of gravitational redshift and Doppler redshift or blueshift. A relation between the observed radiation flux $F_{\text{obs}}$ and the emitted radiation flux $F$
is \cite{Luminet:1979nyg,Kurmanov:2024hpn,Huang:2023ilm,Liu:2024brf}
	\begin{align}
		&F_{\text{obs}}=\frac{F}{{(1+z)}^{4}}.
	\end{align}
Here $(1+z)$ is the redshift factor which is defined as	
	\begin{align}
		1+z=\frac{1+\Omega b \sin\theta_0\sin\alpha}{\sqrt{-g_{tt}-2\Omega g_{t\varphi}-\Omega^{2}g_{\varphi \varphi}}},
	\end{align}
where $\theta_0$ is the inclination angle of the observer and $\alpha$ is the polar angle of an image point on observer's plane \cite{Luminet:1979nyg,Huang:2023ilm}.  
For our discussed quantum corrected black hole, the specific expression for the redshift factor is
	\begin{align}
		1+z=\frac{1+\sqrt{\frac{r^3-\xi^2 \left(r^2-6 r+8\right)}{r^4}} \sin\theta_0 \sin \alpha}{\sqrt{\frac{(r-3) \left(2 \xi^2 (r-2)+r^3\right)}{r^4}}}.
	\end{align} 
Without loss of generality, we shall only work with $\theta_0\approx \frac{\pi}{2}$ in this work, i.e., the observer is located at a small angle above the accretion disk plane. 
Since our main interest is to explore how the correction parameter $\xi$ affects the observed radiation flux, 
 we only focus on the observed fluxes in directions with $\alpha=\frac{\pi}{2}$ and $\alpha=\frac{3\pi}{2}$, which respectively correspond to the least and most redshift directions. 	

In Fig.\ref{flux}, we show three radiation flux cases: the first is the emitted radiation flux by the disk, the second is the observed radiation flux in the direction with $\alpha=\frac{\pi}{2}$, and the third is the observed radiation flux in the direction with $\alpha=\frac{3\pi}{2}$. For each case, we plot several typical radiation flux curves with different values of $\xi$. 
The mass accretion rate is assumed to be a constant which is chosen as $10^6$ here . It is observed that the correction parameter $\xi$ has an apparent effect on the peak of the radiation flux. For relatively large values of $\xi$, the peaks of the radiation fluxes in the first and third cases decrease significantly, compared with the Schwarzschild case. The effect of the correction parameter $\xi$ on the flux in the far region from the black hole is negligible. So, it could be possible to constraint the correction parameter $\xi$ through observations of the peak of the radiation flux. 

	\begin{figure*}[ht]
		\centering
		\subfigure[]{
			\begin{minipage}[t]{0.3\textwidth}
				\centering
				\includegraphics[scale=0.4]{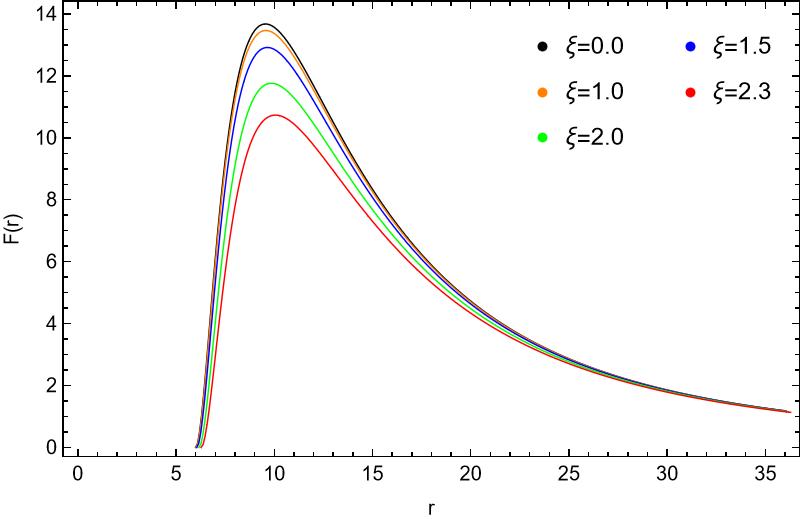}
			\end{minipage}
		}
		\hfill
		\subfigure[]{
			\begin{minipage}[t]{0.3\textwidth}
				\centering
				\includegraphics[scale=0.4]{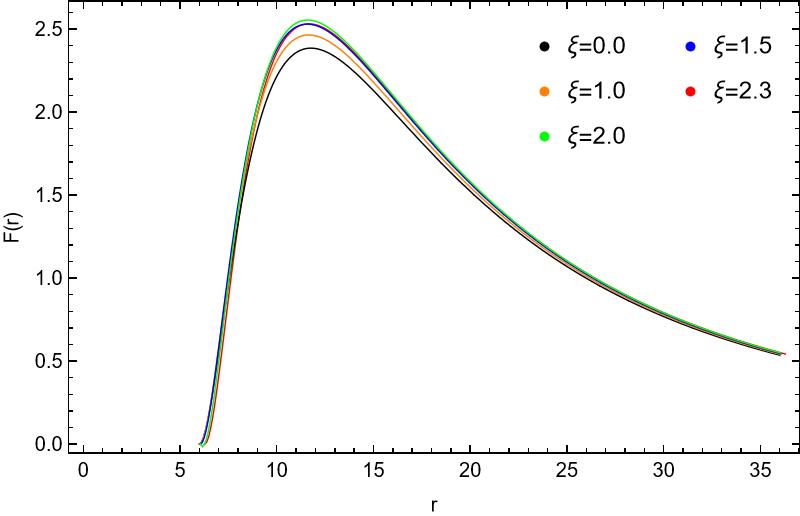}
			\end{minipage}
		}
		\centering
		\hfill
		\subfigure[]{
			\begin{minipage}[t]{0.3\textwidth}
				\centering
				\includegraphics[scale=0.4]{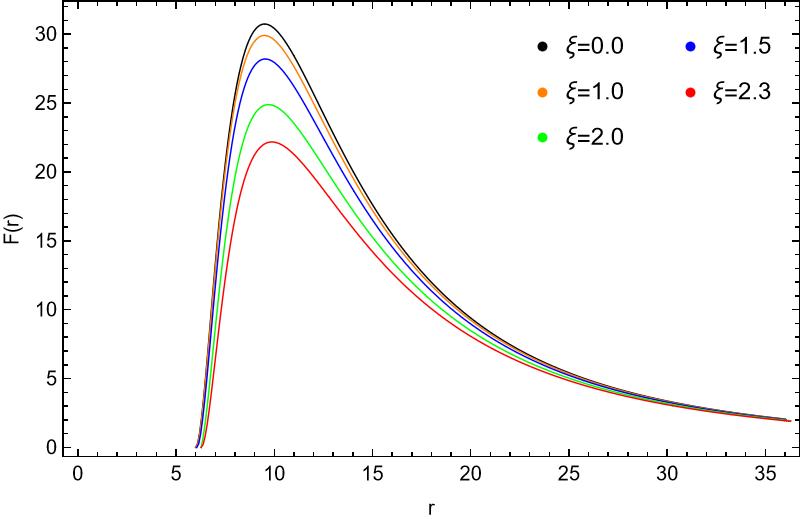}
			\end{minipage}
		}
		\centering
		
		\caption{Radiation fluxes of three cases with different values of $\xi$. Left panel: the radiation flux of the first case. Middle panel: the radiation flux of the second case. Right panel: the radiation flux of the third case.}
		\label{flux}
	\end{figure*}

	\begin{figure*}[htb]
		\centering
		\subfigure[]{
			\begin{minipage}[t]{0.3\textwidth}
				\centering
				\includegraphics[scale=0.48]{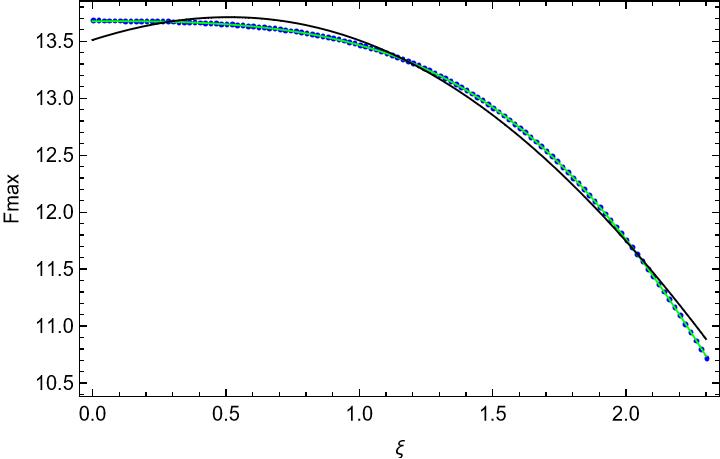}
			\end{minipage}
		}
		\hfill
		\subfigure[]{
			\begin{minipage}[t]{0.3\textwidth}
				\centering
				\includegraphics[scale=0.48]{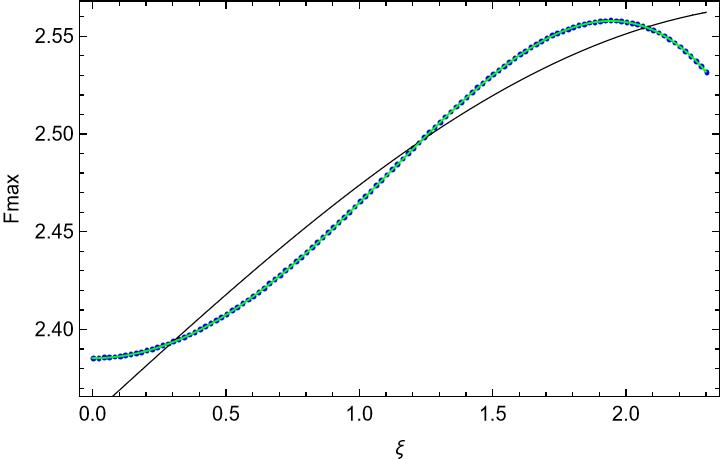}
			\end{minipage}
		}
		\centering
		\hfill
		\subfigure[]{
			\begin{minipage}[t]{0.3\textwidth}
				\centering
				\includegraphics[scale=0.47]{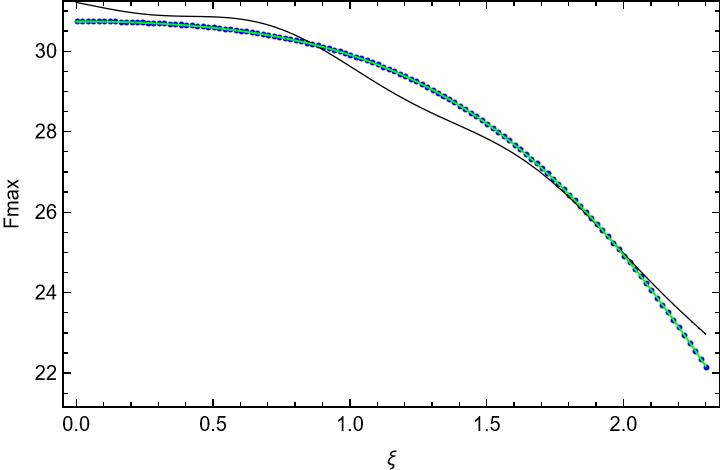}
			\end{minipage}
		}
		\centering
		
		\caption{The fitting of the relation between the maximum of the radiation flux and the parameter $\xi$ for three cases. The solid black points are the numerical data. Fitting with Fourier function (solid curve in black), fitting with polynomial function (solid curve in green). Left panel: fitting for the first case. Middle panel: fitting for the second case. Right panel: fitting for the third case. The fitting curves from exponential function almost exactly overlap with the polynomial ones, so we do not show them here.}
		\label{exp}
	\end{figure*}

	\begin{table}[h]
		\renewcommand{\arraystretch}{2}
		\caption{\label{2}
			MSE of Fourier functions, exponential functions, and polynomial functions with accurate values for the maximum value of the radiative flux $F_{max}$. }
		\begin{ruledtabular}
			\begin{tabular}{cccc}
				Functions&$F$&$F_{\text{obs}}(\alpha=\frac{\pi}{2})$&$F_{\text{obs}}(\alpha=\frac{3\pi}{2})$\\
				\hline
				Fourier&$4.24961\times10^{-3}$&$1.32092\times10^{-3}$&$0.112509$\\

				Exponential&$1.70123\times10^{-6} $& $3.53409\times10^{-8} $ &$7.39967\times10^{-7} $  \\
				
				Polynomial&$5.4641\times10^{-9} $&$1.09207\times10^{-11}$&$3.68347\times10^{-8}$
				
			\end{tabular}
		\end{ruledtabular}
	\end{table}
			\begin{table*}[ht]
			\renewcommand{\arraystretch}{1.5}
			\caption{\label{3}
		 The coefficients of the polynomial fitting functions for the three fluxes.}
			\begin{ruledtabular}
				\begin{tabular}{cccccccc}
					Fluxes&$c_{1}$&$c_{2}$&$c_{3}$&$c_{4}$&$c_{5}$&$c_{6}$&$c_{7}$\\
					\hline
					F&0.00279731&0.0256613&-0.183315&0.06206&-0.122366&0.00538546&13.6779\\

					$F_{\text{obs}}(\alpha=\frac{\pi}{2})$&0.008982 & 0.057642 &-0.433832 &0.149686&-0.626016&0.013344&30.739662\\
					
					$F_{\text{obs}}(\alpha=\frac{3\pi}{2})$& 0.000779&-0.003896& -0.005304& -0.005184 &0.093448&-0.000349&2.385323
					
				\end{tabular}
			\end{ruledtabular}
		\end{table*}

We also explore the relations between the maxima of the radiation fluxes and the correction parameter $\xi$ for the three cases.
The fitting functions used to describe the relations are chosen to be similar as that in \cite{Huang:2023ilm}, which are
\begin{align}
			&F=a_{1}e^{-(\frac{\xi-a_{2}}{a_{3}})^{2}}+a_{4}e^{-(\frac{\xi-a_{5}}{a_{6}})^{2}}\label{24},\\
			&F=b_1+b_2\cos(b_3\xi)+b_4\sin( b_5\xi),\\
			&F=c_1\xi^6+c_2\xi^5+c_3\xi^4+c_4\xi^3+c_5\xi^2+c_6\xi+c_7,\label{26}
		\end{align}
where $a_i,b_i,c_i$ are constant fitting parameters. 
In each case, we numerically solve the maxima of the fluxes for a series of values of $\xi$, which are represented by solid points 
in Fig.~\ref{exp}. Then, we perform fitting with three fitting functions and plot the fitted functions in the same figure. 
It is found that the polynomial function and the exponential function fit the numerical data points almost perfectly in each case,  
however, the Fourier function shows an obvious deviation. 
It is also clear that 
the maxima of the radiation fluxes in the first and third cases are both monotonically decrease with the increase of $\xi$.  
However, in the second case, the maximum of the radiation flux first increase with the increase $\xi$, and then decrease when $\xi$ is roughly greater than 2.

 Mathematically, the mean square error is used to evaluate a fitting function to given data. 		
 In Tab.\ref{2}, we present the mean square errors of the three fitting functions in each case. We can see that the polynomial function has the smallest mean square error. However, both polynomial and exponential functions fit the data well enough. On the contrary, the Fourier function has a much larger mean square error, resulting in the worse fitting. 
 In Tab.\ref{3}, we list the values of fitting parameters in the polynomial function for the three cases.

\section{Summary}
We mainly focus on how the quantum correction parameter $\xi$ affect the properties of the black hole in this work. After calculating the shadow radius of the quantum corrected black hole, we provide a constraint on the correction parameter with the M87* observational data, $0\leq\xi\lesssim2.304$, which is a little better than constraint obtained with Sgr A* \cite{Konoplya:2024lch}. We also calculate the conserved energy, angular momentum, angular velocity for particles on a circular orbit around the black hole. The effects of variation of the correction parameter on these quantities are discussed. It is found that the correction parameter has a significant effect on the angular momenta of particles on the circular orbits even in the far region from the black hole. This results in a possibility to check the parameter $\xi$ through observation of angular momenta of particles. 
The radius of ISCO increases with the increase the parameter $\xi$ and has a visible deviation from Schwarzschild case when $\xi>1$. 
The radiative efficiencies of black holes with different parameters are calculated. We found that the efficiency decreases with the increase of $\xi$.

We also consider the effect of the parameter $\xi$ on the radiation fluxes from the thin disk around the black hole. We study three cases for the radiation fluxes. One is the emitted flux by the disk, the second and third cases are the observed fluxes in directions with $\alpha=\frac{\pi}{2} ~\text{and}~ \frac{3\pi}{2}$ by a far observer located above the disk plane at a small angle. The parameter $\xi$ affects the peak of the fluxes significantly in the first and third cases, while the effect in the second case is relatively small. 
The maxima of the fluxes decrease with the increase of $\xi$ in the first and third cases, while in the second case, the maximum of the flux first increase and then decrease with the increase of $\xi$.  We then examine the relation between the maximum of the flux and the parameter $\xi$ for each case. It is found that the polynomial function fits the numerical data best among our chosen three fitting functions. Furthermore,   
the polynomial function proves to have the smallest mean square error compared with the exponential function and Fourier function. 
Our findings elucidate the effects of quantum correction parameter on several properties of the black hole. They also demonstrate the possible ways to identify the quantum correction parameter.  

\begin{acknowledgments}
This work is partially supported by Guangdong Major Project of Basic and Applied Basic Research (No.2020B0301030008).
\end{acknowledgments}
		
\bibliography{ref}

\begin{thebibliography}{63}%
\makeatletter
\providecommand \@ifxundefined [1]{%
 \@ifx{#1\undefined}
}%
\providecommand \@ifnum [1]{%
 \ifnum #1\expandafter \@firstoftwo
 \else \expandafter \@secondoftwo
 \fi
}%
\providecommand \@ifx [1]{%
 \ifx #1\expandafter \@firstoftwo
 \else \expandafter \@secondoftwo
 \fi
}%
\providecommand \natexlab [1]{#1}%
\providecommand \enquote  [1]{``#1''}%
\providecommand \bibnamefont  [1]{#1}%
\providecommand \bibfnamefont [1]{#1}%
\providecommand \citenamefont [1]{#1}%
\providecommand \href@noop [0]{\@secondoftwo}%
\providecommand \href [0]{\begingroup \@sanitize@url \@href}%
\providecommand \@href[1]{\@@startlink{#1}\@@href}%
\providecommand \@@href[1]{\endgroup#1\@@endlink}%
\providecommand \@sanitize@url [0]{\catcode `\\12\catcode `\$12\catcode
  `\&12\catcode `\#12\catcode `\^12\catcode `\_12\catcode `\%12\relax}%
\providecommand \@@startlink[1]{}%
\providecommand \@@endlink[0]{}%
\providecommand \url  [0]{\begingroup\@sanitize@url \@url }%
\providecommand \@url [1]{\endgroup\@href {#1}{\urlprefix }}%
\providecommand \urlprefix  [0]{URL }%
\providecommand \Eprint [0]{\href }%
\providecommand \doibase [0]{https://doi.org/}%
\providecommand \selectlanguage [0]{\@gobble}%
\providecommand \bibinfo  [0]{\@secondoftwo}%
\providecommand \bibfield  [0]{\@secondoftwo}%
\providecommand \translation [1]{[#1]}%
\providecommand \BibitemOpen [0]{}%
\providecommand \bibitemStop [0]{}%
\providecommand \bibitemNoStop [0]{.\EOS\space}%
\providecommand \EOS [0]{\spacefactor3000\relax}%
\providecommand \BibitemShut  [1]{\csname bibitem#1\endcsname}%
\let\auto@bib@innerbib\@empty
\bibitem [{\citenamefont {Penrose}(1965)}]{Penrose:1964wq}%
  \BibitemOpen
  \bibfield  {author} {\bibinfo {author} {\bibfnamefont {R.}~\bibnamefont
  {Penrose}},\ }\bibfield  {title} {\bibinfo {title} {{Gravitational collapse
  and space-time singularities}},\ }\href
  {https://doi.org/10.1103/PhysRevLett.14.57} {\bibfield  {journal} {\bibinfo
  {journal} {Phys. Rev. Lett.}\ }\textbf {\bibinfo {volume} {14}},\ \bibinfo
  {pages} {57} (\bibinfo {year} {1965})}\BibitemShut {NoStop}%
\bibitem [{\citenamefont {Hawking}\ and\ \citenamefont
  {Penrose}(1970)}]{Hawking:1970zqf}%
  \BibitemOpen
  \bibfield  {author} {\bibinfo {author} {\bibfnamefont {S.~W.}\ \bibnamefont
  {Hawking}}\ and\ \bibinfo {author} {\bibfnamefont {R.}~\bibnamefont
  {Penrose}},\ }\bibfield  {title} {\bibinfo {title} {{The Singularities of
  gravitational collapse and cosmology}},\ }\href
  {https://doi.org/10.1098/rspa.1970.0021} {\bibfield  {journal} {\bibinfo
  {journal} {Proc. Roy. Soc. Lond. A}\ }\textbf {\bibinfo {volume} {314}},\
  \bibinfo {pages} {529} (\bibinfo {year} {1970})}\BibitemShut {NoStop}%
\bibitem [{\citenamefont {Marolf}(2017)}]{Marolf:2017jkr}%
  \BibitemOpen
  \bibfield  {author} {\bibinfo {author} {\bibfnamefont {D.}~\bibnamefont
  {Marolf}},\ }\bibfield  {title} {\bibinfo {title} {{The Black Hole
  information problem: past, present, and future}},\ }\href
  {https://doi.org/10.1088/1361-6633/aa77cc} {\bibfield  {journal} {\bibinfo
  {journal} {Rept. Prog. Phys.}\ }\textbf {\bibinfo {volume} {80}},\ \bibinfo
  {pages} {092001} (\bibinfo {year} {2017})},\ \Eprint
  {https://arxiv.org/abs/1703.02143} {arXiv:1703.02143 [gr-qc]} \BibitemShut
  {NoStop}%
\bibitem [{\citenamefont {Afshordi}\ \emph {et~al.}(2024)\citenamefont
  {Afshordi} \emph {et~al.}}]{Buoninfante:2024oxl}%
  \BibitemOpen
  \bibfield  {author} {\bibinfo {author} {\bibfnamefont {N.}~\bibnamefont
  {Afshordi}} \emph {et~al.},\ }\bibfield  {title} {\bibinfo {title} {{Black
  Holes Inside and Out 2024: visions for the future of black hole physics}}\
  }(\bibinfo {year} {2024})\ \Eprint {https://arxiv.org/abs/2410.14414}
  {arXiv:2410.14414 [gr-qc]} \BibitemShut {NoStop}%
\bibitem [{\citenamefont {Thiemann}(2007)}]{Thiemann:2007pyv}%
  \BibitemOpen
  \bibfield  {author} {\bibinfo {author} {\bibfnamefont {T.}~\bibnamefont
  {Thiemann}},\ }\href {https://doi.org/10.1017/CBO9780511755682} {\emph
  {\bibinfo {title} {{Modern Canonical Quantum General Relativity}}}},\
  Cambridge Monographs on Mathematical Physics\ (\bibinfo  {publisher}
  {Cambridge University Press},\ \bibinfo {year} {2007})\BibitemShut {NoStop}%
\bibitem [{\citenamefont {Ashtekar}\ and\ \citenamefont
  {Lewandowski}(2004)}]{Ashtekar:2004eh}%
  \BibitemOpen
  \bibfield  {author} {\bibinfo {author} {\bibfnamefont {A.}~\bibnamefont
  {Ashtekar}}\ and\ \bibinfo {author} {\bibfnamefont {J.}~\bibnamefont
  {Lewandowski}},\ }\bibfield  {title} {\bibinfo {title} {{Background
  independent quantum gravity: A Status report}},\ }\href
  {https://doi.org/10.1088/0264-9381/21/15/R01} {\bibfield  {journal} {\bibinfo
   {journal} {Class. Quant. Grav.}\ }\textbf {\bibinfo {volume} {21}},\
  \bibinfo {pages} {R53} (\bibinfo {year} {2004})},\ \Eprint
  {https://arxiv.org/abs/gr-qc/0404018} {arXiv:gr-qc/0404018} \BibitemShut
  {NoStop}%
\bibitem [{\citenamefont {Addazi}\ \emph {et~al.}(2022)\citenamefont {Addazi}
  \emph {et~al.}}]{Addazi2022}%
  \BibitemOpen
  \bibfield  {author} {\bibinfo {author} {\bibfnamefont {A.}~\bibnamefont
  {Addazi}} \emph {et~al.},\ }\bibfield  {title} {\bibinfo {title} {{Quantum
  gravity phenomenology at the dawn of the multi-messenger era\textemdash{}A
  review}},\ }\href {https://doi.org/10.1016/j.ppnp.2022.103948} {\bibfield
  {journal} {\bibinfo  {journal} {Prog. Part. Nucl. Phys.}\ }\textbf {\bibinfo
  {volume} {125}},\ \bibinfo {pages} {103948} (\bibinfo {year} {2022})},\
  \Eprint {https://arxiv.org/abs/2111.05659} {arXiv:2111.05659 [hep-ph]}
  \BibitemShut {NoStop}%
\bibitem [{\citenamefont {Zhang}\ \emph {et~al.}(2024)\citenamefont {Zhang},
  \citenamefont {Lewandowski}, \citenamefont {Ma},\ and\ \citenamefont
  {Yang}}]{Zhang:2024khj}%
  \BibitemOpen
  \bibfield  {author} {\bibinfo {author} {\bibfnamefont {C.}~\bibnamefont
  {Zhang}}, \bibinfo {author} {\bibfnamefont {J.}~\bibnamefont {Lewandowski}},
  \bibinfo {author} {\bibfnamefont {Y.}~\bibnamefont {Ma}},\ and\ \bibinfo
  {author} {\bibfnamefont {J.}~\bibnamefont {Yang}},\ }\bibfield  {title}
  {\bibinfo {title} {{Black Holes and Covariance in Effective Quantum
  Gravity}},\ }\href@noop {} {\  (\bibinfo {year} {2024})},\ \Eprint
  {https://arxiv.org/abs/2407.10168} {arXiv:2407.10168 [gr-qc]} \BibitemShut
  {NoStop}%
\bibitem [{\citenamefont {Lewandowski}\ \emph {et~al.}(2023)\citenamefont
  {Lewandowski}, \citenamefont {Ma}, \citenamefont {Yang},\ and\ \citenamefont
  {Zhang}}]{Lewandowski:2022zce}%
  \BibitemOpen
  \bibfield  {author} {\bibinfo {author} {\bibfnamefont {J.}~\bibnamefont
  {Lewandowski}}, \bibinfo {author} {\bibfnamefont {Y.}~\bibnamefont {Ma}},
  \bibinfo {author} {\bibfnamefont {J.}~\bibnamefont {Yang}},\ and\ \bibinfo
  {author} {\bibfnamefont {C.}~\bibnamefont {Zhang}},\ }\bibfield  {title}
  {\bibinfo {title} {{Quantum Oppenheimer-Snyder and Swiss Cheese Models}},\
  }\href {https://doi.org/10.1103/PhysRevLett.130.101501} {\bibfield  {journal}
  {\bibinfo  {journal} {Phys. Rev. Lett.}\ }\textbf {\bibinfo {volume} {130}},\
  \bibinfo {pages} {101501} (\bibinfo {year} {2023})},\ \Eprint
  {https://arxiv.org/abs/2210.02253} {arXiv:2210.02253 [gr-qc]} \BibitemShut
  {NoStop}%
\bibitem [{\citenamefont {Kelly}\ \emph {et~al.}(2020)\citenamefont {Kelly},
  \citenamefont {Santacruz},\ and\ \citenamefont
  {Wilson-Ewing}}]{Kelly:2020lec}%
  \BibitemOpen
  \bibfield  {author} {\bibinfo {author} {\bibfnamefont {J.~G.}\ \bibnamefont
  {Kelly}}, \bibinfo {author} {\bibfnamefont {R.}~\bibnamefont {Santacruz}},\
  and\ \bibinfo {author} {\bibfnamefont {E.}~\bibnamefont {Wilson-Ewing}},\
  }\bibfield  {title} {\bibinfo {title} {{Effective loop quantum gravity
  framework for vacuum spherically symmetric spacetimes}},\ }\href
  {https://doi.org/10.1103/PhysRevD.102.106024} {\bibfield  {journal} {\bibinfo
   {journal} {Phys. Rev. D}\ }\textbf {\bibinfo {volume} {102}},\ \bibinfo
  {pages} {106024} (\bibinfo {year} {2020})},\ \Eprint
  {https://arxiv.org/abs/2006.09302} {arXiv:2006.09302 [gr-qc]} \BibitemShut
  {NoStop}%
\bibitem [{\citenamefont {Yang}\ \emph {et~al.}(2023)\citenamefont {Yang},
  \citenamefont {Zhang},\ and\ \citenamefont {Ma}}]{Yang:2022btw}%
  \BibitemOpen
  \bibfield  {author} {\bibinfo {author} {\bibfnamefont {J.}~\bibnamefont
  {Yang}}, \bibinfo {author} {\bibfnamefont {C.}~\bibnamefont {Zhang}},\ and\
  \bibinfo {author} {\bibfnamefont {Y.}~\bibnamefont {Ma}},\ }\bibfield
  {title} {\bibinfo {title} {{Shadow and stability of quantum-corrected black
  holes}},\ }\href {https://doi.org/10.1140/epjc/s10052-023-11800-8} {\bibfield
   {journal} {\bibinfo  {journal} {Eur. Phys. J. C}\ }\textbf {\bibinfo
  {volume} {83}},\ \bibinfo {pages} {619} (\bibinfo {year} {2023})},\ \Eprint
  {https://arxiv.org/abs/2211.04263} {arXiv:2211.04263 [gr-qc]} \BibitemShut
  {NoStop}%
\bibitem [{\citenamefont {Cao}\ \emph {et~al.}(2024)\citenamefont {Cao},
  \citenamefont {Chen}, \citenamefont {Wu}, \citenamefont {Xie},\ and\
  \citenamefont {Zhou}}]{Cao:2024oud}%
  \BibitemOpen
  \bibfield  {author} {\bibinfo {author} {\bibfnamefont {L.-M.}\ \bibnamefont
  {Cao}}, \bibinfo {author} {\bibfnamefont {J.-N.}\ \bibnamefont {Chen}},
  \bibinfo {author} {\bibfnamefont {L.-B.}\ \bibnamefont {Wu}}, \bibinfo
  {author} {\bibfnamefont {L.}~\bibnamefont {Xie}},\ and\ \bibinfo {author}
  {\bibfnamefont {Y.-S.}\ \bibnamefont {Zhou}},\ }\bibfield  {title} {\bibinfo
  {title} {{The pseudospectrum and spectrum (in)stability of quantum corrected
  Schwarzschild black hole}},\ }\href
  {https://doi.org/10.1007/s11433-024-2435-5} {\bibfield  {journal} {\bibinfo
  {journal} {Sci. China Phys. Mech. Astron.}\ }\textbf {\bibinfo {volume}
  {67}},\ \bibinfo {pages} {100412} (\bibinfo {year} {2024})},\ \Eprint
  {https://arxiv.org/abs/2401.09907} {arXiv:2401.09907 [gr-qc]} \BibitemShut
  {NoStop}%
\bibitem [{\citenamefont {Zhang}\ and\ \citenamefont
  {Wang}(2024)}]{Zhang:2024svj}%
  \BibitemOpen
  \bibfield  {author} {\bibinfo {author} {\bibfnamefont {C.}~\bibnamefont
  {Zhang}}\ and\ \bibinfo {author} {\bibfnamefont {A.}~\bibnamefont {Wang}},\
  }\bibfield  {title} {\bibinfo {title} {{Quasi-normal modes of loop quantum
  black holes formed from gravitational collapse}},\ }\href
  {https://doi.org/10.1088/1475-7516/2024/10/070} {\bibfield  {journal}
  {\bibinfo  {journal} {JCAP}\ }\textbf {\bibinfo {volume} {10}},\ \bibinfo
  {pages} {070}},\ \Eprint {https://arxiv.org/abs/2407.19654} {arXiv:2407.19654
  [gr-qc]} \BibitemShut {NoStop}%
\bibitem [{\citenamefont
  {Skvortsova}(2024{\natexlab{a}})}]{Skvortsova:2024atk}%
  \BibitemOpen
  \bibfield  {author} {\bibinfo {author} {\bibfnamefont {M.}~\bibnamefont
  {Skvortsova}},\ }\bibfield  {title} {\bibinfo {title} {{Quasinormal
  Frequencies of Fields with Various Spin in the Quantum
  Oppenheimer\textendash{}Snyder Model of Black Holes}},\ }\href
  {https://doi.org/10.1002/prop.202400132} {\bibfield  {journal} {\bibinfo
  {journal} {Fortsch. Phys.}\ }\textbf {\bibinfo {volume} {72}},\ \bibinfo
  {pages} {2400132} (\bibinfo {year} {2024}{\natexlab{a}})},\ \Eprint
  {https://arxiv.org/abs/2405.06390} {arXiv:2405.06390 [gr-qc]} \BibitemShut
  {NoStop}%
\bibitem [{\citenamefont {Zinhailo}(2024)}]{Zinhailo:2024kbq}%
  \BibitemOpen
  \bibfield  {author} {\bibinfo {author} {\bibfnamefont {A.~F.}\ \bibnamefont
  {Zinhailo}},\ }\bibfield  {title} {\bibinfo {title} {{Black Hole in the
  Quantum Oppenheimer-Snyder model: long lived modes and the overtones'
  behavior}}\ }\href {https://doi.org/10.13140/RG.2.2.26785.01124}
  {10.13140/RG.2.2.26785.01124} (\bibinfo {year} {2024})\BibitemShut {NoStop}%
\bibitem [{\citenamefont {Gong}\ \emph {et~al.}(2024)\citenamefont {Gong},
  \citenamefont {Li}, \citenamefont {Zhang}, \citenamefont {Fu},\ and\
  \citenamefont {Wu}}]{Gong:2023ghh}%
  \BibitemOpen
  \bibfield  {author} {\bibinfo {author} {\bibfnamefont {H.}~\bibnamefont
  {Gong}}, \bibinfo {author} {\bibfnamefont {S.}~\bibnamefont {Li}}, \bibinfo
  {author} {\bibfnamefont {D.}~\bibnamefont {Zhang}}, \bibinfo {author}
  {\bibfnamefont {G.}~\bibnamefont {Fu}},\ and\ \bibinfo {author}
  {\bibfnamefont {J.-P.}\ \bibnamefont {Wu}},\ }\bibfield  {title} {\bibinfo
  {title} {{Quasinormal modes of quantum-corrected black holes}},\ }\href
  {https://doi.org/10.1103/PhysRevD.110.044040} {\bibfield  {journal} {\bibinfo
   {journal} {Phys. Rev. D}\ }\textbf {\bibinfo {volume} {110}},\ \bibinfo
  {pages} {044040} (\bibinfo {year} {2024})},\ \Eprint
  {https://arxiv.org/abs/2312.17639} {arXiv:2312.17639 [gr-qc]} \BibitemShut
  {NoStop}%
\bibitem [{\citenamefont {Malik}(2024)}]{Malik:2024nhy}%
  \BibitemOpen
  \bibfield  {author} {\bibinfo {author} {\bibfnamefont {Z.}~\bibnamefont
  {Malik}},\ }\bibfield  {title} {\bibinfo {title} {{Perturbations and
  Quasinormal Modes of the Dirac Field in Effective Quantum Gravity}},\
  }\href@noop {} {\  (\bibinfo {year} {2024})},\ \Eprint
  {https://arxiv.org/abs/2409.01561} {arXiv:2409.01561 [gr-qc]} \BibitemShut
  {NoStop}%
\bibitem [{\citenamefont {Konoplya}\ and\ \citenamefont
  {Stashko}(2024)}]{Konoplya:2024lch}%
  \BibitemOpen
  \bibfield  {author} {\bibinfo {author} {\bibfnamefont {R.~A.}\ \bibnamefont
  {Konoplya}}\ and\ \bibinfo {author} {\bibfnamefont {O.~S.}\ \bibnamefont
  {Stashko}},\ }\bibfield  {title} {\bibinfo {title} {{Probing the Effective
  Quantum Gravity via Quasinormal Modes and Shadows of Black Holes}},\
  }\href@noop {} {\  (\bibinfo {year} {2024})},\ \Eprint
  {https://arxiv.org/abs/2408.02578} {arXiv:2408.02578 [gr-qc]} \BibitemShut
  {NoStop}%
\bibitem [{\citenamefont
  {Skvortsova}(2024{\natexlab{b}})}]{Skvortsova:2024msa}%
  \BibitemOpen
  \bibfield  {author} {\bibinfo {author} {\bibfnamefont {M.}~\bibnamefont
  {Skvortsova}},\ }\bibfield  {title} {\bibinfo {title} {{Quantum corrected
  black holes: testing the correspondence between grey-body factors and
  quasinormal modes}},\ }\href@noop {} {\  (\bibinfo {year}
  {2024}{\natexlab{b}})},\ \Eprint {https://arxiv.org/abs/2411.06007}
  {arXiv:2411.06007 [gr-qc]} \BibitemShut {NoStop}%
\bibitem [{\citenamefont {Zhang}\ \emph {et~al.}(2023)\citenamefont {Zhang},
  \citenamefont {Ma},\ and\ \citenamefont {Yang}}]{Zhang:2023okw}%
  \BibitemOpen
  \bibfield  {author} {\bibinfo {author} {\bibfnamefont {C.}~\bibnamefont
  {Zhang}}, \bibinfo {author} {\bibfnamefont {Y.}~\bibnamefont {Ma}},\ and\
  \bibinfo {author} {\bibfnamefont {J.}~\bibnamefont {Yang}},\ }\bibfield
  {title} {\bibinfo {title} {{Black hole image encoding quantum gravity
  information}},\ }\href {https://doi.org/10.1103/PhysRevD.108.104004}
  {\bibfield  {journal} {\bibinfo  {journal} {Phys. Rev. D}\ }\textbf {\bibinfo
  {volume} {108}},\ \bibinfo {pages} {104004} (\bibinfo {year} {2023})},\
  \Eprint {https://arxiv.org/abs/2302.02800} {arXiv:2302.02800 [gr-qc]}
  \BibitemShut {NoStop}%
\bibitem [{\citenamefont {Ye}\ \emph {et~al.}(2024)\citenamefont {Ye},
  \citenamefont {He}, \citenamefont {Zhou}, \citenamefont {Huang},\ and\
  \citenamefont {Huang}}]{Ye:2023qks}%
  \BibitemOpen
  \bibfield  {author} {\bibinfo {author} {\bibfnamefont {J.-P.}\ \bibnamefont
  {Ye}}, \bibinfo {author} {\bibfnamefont {Z.-Q.}\ \bibnamefont {He}}, \bibinfo
  {author} {\bibfnamefont {A.-X.}\ \bibnamefont {Zhou}}, \bibinfo {author}
  {\bibfnamefont {Z.-Y.}\ \bibnamefont {Huang}},\ and\ \bibinfo {author}
  {\bibfnamefont {J.-H.}\ \bibnamefont {Huang}},\ }\bibfield  {title} {\bibinfo
  {title} {{Shadows and photon rings of a quantum black hole}},\ }\href
  {https://doi.org/10.1016/j.physletb.2024.138566} {\bibfield  {journal}
  {\bibinfo  {journal} {Phys. Lett. B}\ }\textbf {\bibinfo {volume} {851}},\
  \bibinfo {pages} {138566} (\bibinfo {year} {2024})},\ \Eprint
  {https://arxiv.org/abs/2312.17724} {arXiv:2312.17724 [gr-qc]} \BibitemShut
  {NoStop}%
\bibitem [{\citenamefont {Liu}\ \emph {et~al.}(2024{\natexlab{a}})\citenamefont
  {Liu}, \citenamefont {Wu},\ and\ \citenamefont {Wang}}]{Liu:2024soc}%
  \BibitemOpen
  \bibfield  {author} {\bibinfo {author} {\bibfnamefont {W.}~\bibnamefont
  {Liu}}, \bibinfo {author} {\bibfnamefont {D.}~\bibnamefont {Wu}},\ and\
  \bibinfo {author} {\bibfnamefont {J.}~\bibnamefont {Wang}},\ }\bibfield
  {title} {\bibinfo {title} {{Light rings and shadows of static black holes in
  effective quantum gravity}},\ }\href
  {https://doi.org/10.1016/j.physletb.2024.139052} {\bibfield  {journal}
  {\bibinfo  {journal} {Phys. Lett. B}\ }\textbf {\bibinfo {volume} {858}},\
  \bibinfo {pages} {139052} (\bibinfo {year} {2024}{\natexlab{a}})},\ \Eprint
  {https://arxiv.org/abs/2408.05569} {arXiv:2408.05569 [gr-qc]} \BibitemShut
  {NoStop}%
\bibitem [{\citenamefont {Peng}\ \emph {et~al.}(2021)\citenamefont {Peng},
  \citenamefont {Guo},\ and\ \citenamefont {Feng}}]{Peng:2020wun}%
  \BibitemOpen
  \bibfield  {author} {\bibinfo {author} {\bibfnamefont {J.}~\bibnamefont
  {Peng}}, \bibinfo {author} {\bibfnamefont {M.}~\bibnamefont {Guo}},\ and\
  \bibinfo {author} {\bibfnamefont {X.-H.}\ \bibnamefont {Feng}},\ }\bibfield
  {title} {\bibinfo {title} {{Influence of quantum correction on black hole
  shadows, photon rings, and lensing rings}},\ }\href
  {https://doi.org/10.1088/1674-1137/ac06bb} {\bibfield  {journal} {\bibinfo
  {journal} {Chin. Phys. C}\ }\textbf {\bibinfo {volume} {45}},\ \bibinfo
  {pages} {085103} (\bibinfo {year} {2021})},\ \Eprint
  {https://arxiv.org/abs/2008.00657} {arXiv:2008.00657 [gr-qc]} \BibitemShut
  {NoStop}%
\bibitem [{\citenamefont {Zhao}\ \emph {et~al.}(2024)\citenamefont {Zhao},
  \citenamefont {Tang},\ and\ \citenamefont {Xu}}]{Zhao:2024elr}%
  \BibitemOpen
  \bibfield  {author} {\bibinfo {author} {\bibfnamefont {L.}~\bibnamefont
  {Zhao}}, \bibinfo {author} {\bibfnamefont {M.}~\bibnamefont {Tang}},\ and\
  \bibinfo {author} {\bibfnamefont {Z.}~\bibnamefont {Xu}},\ }\bibfield
  {title} {\bibinfo {title} {{The lensing effect of quantum-corrected black
  hole and parameter constraints from EHT observations}},\ }\href
  {https://doi.org/10.1140/epjc/s10052-024-13342-z} {\bibfield  {journal}
  {\bibinfo  {journal} {Eur. Phys. J. C}\ }\textbf {\bibinfo {volume} {84}},\
  \bibinfo {pages} {971} (\bibinfo {year} {2024})},\ \Eprint
  {https://arxiv.org/abs/2403.18606} {arXiv:2403.18606 [gr-qc]} \BibitemShut
  {NoStop}%
\bibitem [{\citenamefont {Zi}\ and\ \citenamefont {Kumar}(2024)}]{Zi:2024jla}%
  \BibitemOpen
  \bibfield  {author} {\bibinfo {author} {\bibfnamefont {T.}~\bibnamefont
  {Zi}}\ and\ \bibinfo {author} {\bibfnamefont {S.}~\bibnamefont {Kumar}},\
  }\bibfield  {title} {\bibinfo {title} {{Eccentric extreme mass-ratio
  inspirals: A gateway to probe quantum gravity effects}},\ }\href@noop {} {\
  (\bibinfo {year} {2024})},\ \Eprint {https://arxiv.org/abs/2409.17765}
  {arXiv:2409.17765 [gr-qc]} \BibitemShut {NoStop}%
\bibitem [{\citenamefont {You}\ \emph {et~al.}(2024)\citenamefont {You},
  \citenamefont {Feng}, \citenamefont {Wang}, \citenamefont {Hu},\ and\
  \citenamefont {Deng}}]{You:2024jeu}%
  \BibitemOpen
  \bibfield  {author} {\bibinfo {author} {\bibfnamefont {L.}~\bibnamefont
  {You}}, \bibinfo {author} {\bibfnamefont {Y.-H.}\ \bibnamefont {Feng}},
  \bibinfo {author} {\bibfnamefont {R.-B.}\ \bibnamefont {Wang}}, \bibinfo
  {author} {\bibfnamefont {X.-R.}\ \bibnamefont {Hu}},\ and\ \bibinfo {author}
  {\bibfnamefont {J.-B.}\ \bibnamefont {Deng}},\ }\bibfield  {title} {\bibinfo
  {title} {{Decoding Quantum Gravity Information with Black Hole Accretion
  Disk}},\ }\href {https://doi.org/10.3390/universe10100393} {\bibfield
  {journal} {\bibinfo  {journal} {Universe}\ }\textbf {\bibinfo {volume}
  {10}},\ \bibinfo {pages} {393} (\bibinfo {year} {2024})},\ \Eprint
  {https://arxiv.org/abs/2404.01418} {arXiv:2404.01418 [gr-qc]} \BibitemShut
  {NoStop}%
\bibitem [{\citenamefont {Liu}\ \emph {et~al.}(2024{\natexlab{b}})\citenamefont
  {Liu}, \citenamefont {Lai}, \citenamefont {Pan}, \citenamefont {Huang},\ and\
  \citenamefont {Zou}}]{Liu:2024wal}%
  \BibitemOpen
  \bibfield  {author} {\bibinfo {author} {\bibfnamefont {H.}~\bibnamefont
  {Liu}}, \bibinfo {author} {\bibfnamefont {M.-Y.}\ \bibnamefont {Lai}},
  \bibinfo {author} {\bibfnamefont {X.-Y.}\ \bibnamefont {Pan}}, \bibinfo
  {author} {\bibfnamefont {H.}~\bibnamefont {Huang}},\ and\ \bibinfo {author}
  {\bibfnamefont {D.-C.}\ \bibnamefont {Zou}},\ }\bibfield  {title} {\bibinfo
  {title} {{Gravitational lensing effect of black holes in effective quantum
  gravity}},\ }\href {https://doi.org/10.1103/PhysRevD.110.104039} {\bibfield
  {journal} {\bibinfo  {journal} {Phys. Rev. D}\ }\textbf {\bibinfo {volume}
  {110}},\ \bibinfo {pages} {104039} (\bibinfo {year} {2024}{\natexlab{b}})},\
  \Eprint {https://arxiv.org/abs/2408.11603} {arXiv:2408.11603 [gr-qc]}
  \BibitemShut {NoStop}%
\bibitem [{\citenamefont {Li}\ and\ \citenamefont {Zhang}(2024)}]{Li:2024afr}%
  \BibitemOpen
  \bibfield  {author} {\bibinfo {author} {\bibfnamefont {H.}~\bibnamefont
  {Li}}\ and\ \bibinfo {author} {\bibfnamefont {X.}~\bibnamefont {Zhang}},\
  }\bibfield  {title} {\bibinfo {title} {{Gravitational Lensing Effects from
  Models of Loop Quantum Gravity with Rigorous Quantum Parameters}}\ }\href
  {https://doi.org/10.20944/preprints202409.1122.v1}
  {10.20944/preprints202409.1122.v1} (\bibinfo {year} {2024})\BibitemShut
  {NoStop}%
\bibitem [{\citenamefont {Shakura}\ and\ \citenamefont
  {Sunyaev}(1973)}]{Shakura:1972te}%
  \BibitemOpen
  \bibfield  {author} {\bibinfo {author} {\bibfnamefont {N.~I.}\ \bibnamefont
  {Shakura}}\ and\ \bibinfo {author} {\bibfnamefont {R.~A.}\ \bibnamefont
  {Sunyaev}},\ }\bibfield  {title} {\bibinfo {title} {{Black holes in binary
  systems. Observational appearance}},\ }\href@noop {} {\bibfield  {journal}
  {\bibinfo  {journal} {Astron. Astrophys.}\ }\textbf {\bibinfo {volume}
  {24}},\ \bibinfo {pages} {337} (\bibinfo {year} {1973})}\BibitemShut
  {NoStop}%
\bibitem [{\citenamefont {Page}\ and\ \citenamefont
  {Thorne}(1974)}]{Page:1974he}%
  \BibitemOpen
  \bibfield  {author} {\bibinfo {author} {\bibfnamefont {D.~N.}\ \bibnamefont
  {Page}}\ and\ \bibinfo {author} {\bibfnamefont {K.~S.}\ \bibnamefont
  {Thorne}},\ }\bibfield  {title} {\bibinfo {title} {{Disk-Accretion onto a
  Black Hole. Time-Averaged Structure of Accretion Disk}},\ }\href
  {https://doi.org/10.1086/152990} {\bibfield  {journal} {\bibinfo  {journal}
  {Astrophys. J.}\ }\textbf {\bibinfo {volume} {191}},\ \bibinfo {pages} {499}
  (\bibinfo {year} {1974})}\BibitemShut {NoStop}%
\bibitem [{\citenamefont {Thorne}(1974)}]{Thorne:1974ve}%
  \BibitemOpen
  \bibfield  {author} {\bibinfo {author} {\bibfnamefont {K.~S.}\ \bibnamefont
  {Thorne}},\ }\bibfield  {title} {\bibinfo {title} {{Disk accretion onto a
  black hole. 2. Evolution of the hole.}},\ }\href
  {https://doi.org/10.1086/152991} {\bibfield  {journal} {\bibinfo  {journal}
  {Astrophys. J.}\ }\textbf {\bibinfo {volume} {191}},\ \bibinfo {pages} {507}
  (\bibinfo {year} {1974})}\BibitemShut {NoStop}%
\bibitem [{\citenamefont {Kong}\ \emph {et~al.}(2014)\citenamefont {Kong},
  \citenamefont {Li},\ and\ \citenamefont {Bambi}}]{Kong:2014wha}%
  \BibitemOpen
  \bibfield  {author} {\bibinfo {author} {\bibfnamefont {L.}~\bibnamefont
  {Kong}}, \bibinfo {author} {\bibfnamefont {Z.}~\bibnamefont {Li}},\ and\
  \bibinfo {author} {\bibfnamefont {C.}~\bibnamefont {Bambi}},\ }\bibfield
  {title} {\bibinfo {title} {{Constraints on the spacetime geometry around 10
  stellar-mass black hole candidates from the disk's thermal spectrum}},\
  }\href {https://doi.org/10.1088/0004-637X/797/2/78} {\bibfield  {journal}
  {\bibinfo  {journal} {Astrophys. J.}\ }\textbf {\bibinfo {volume} {797}},\
  \bibinfo {pages} {78} (\bibinfo {year} {2014})},\ \Eprint
  {https://arxiv.org/abs/1405.1508} {arXiv:1405.1508 [gr-qc]} \BibitemShut
  {NoStop}%
\bibitem [{\citenamefont {Pun}\ \emph {et~al.}(2008{\natexlab{a}})\citenamefont
  {Pun}, \citenamefont {Kovacs},\ and\ \citenamefont {Harko}}]{Pun:2008ua}%
  \BibitemOpen
  \bibfield  {author} {\bibinfo {author} {\bibfnamefont {C.~S.~J.}\
  \bibnamefont {Pun}}, \bibinfo {author} {\bibfnamefont {Z.}~\bibnamefont
  {Kovacs}},\ and\ \bibinfo {author} {\bibfnamefont {T.}~\bibnamefont
  {Harko}},\ }\bibfield  {title} {\bibinfo {title} {{Thin accretion disks onto
  brane world black holes}},\ }\href
  {https://doi.org/10.1103/PhysRevD.78.084015} {\bibfield  {journal} {\bibinfo
  {journal} {Phys. Rev. D}\ }\textbf {\bibinfo {volume} {78}},\ \bibinfo
  {pages} {084015} (\bibinfo {year} {2008}{\natexlab{a}})},\ \Eprint
  {https://arxiv.org/abs/0809.1284} {arXiv:0809.1284 [gr-qc]} \BibitemShut
  {NoStop}%
\bibitem [{\citenamefont {Harko}\ \emph {et~al.}(2009)\citenamefont {Harko},
  \citenamefont {Kovacs},\ and\ \citenamefont {Lobo}}]{Harko:2009rp}%
  \BibitemOpen
  \bibfield  {author} {\bibinfo {author} {\bibfnamefont {T.}~\bibnamefont
  {Harko}}, \bibinfo {author} {\bibfnamefont {Z.}~\bibnamefont {Kovacs}},\ and\
  \bibinfo {author} {\bibfnamefont {F.~S.~N.}\ \bibnamefont {Lobo}},\
  }\bibfield  {title} {\bibinfo {title} {{Testing Ho\v{r}ava-Lifshitz gravity
  using thin accretion disk properties}},\ }\href
  {https://doi.org/10.1103/PhysRevD.80.044021} {\bibfield  {journal} {\bibinfo
  {journal} {Phys. Rev. D}\ }\textbf {\bibinfo {volume} {80}},\ \bibinfo
  {pages} {044021} (\bibinfo {year} {2009})},\ \Eprint
  {https://arxiv.org/abs/0907.1449} {arXiv:0907.1449 [gr-qc]} \BibitemShut
  {NoStop}%
\bibitem [{\citenamefont {Harko}\ \emph {et~al.}(2010)\citenamefont {Harko},
  \citenamefont {Kovacs},\ and\ \citenamefont {Lobo}}]{Harko:2009kj}%
  \BibitemOpen
  \bibfield  {author} {\bibinfo {author} {\bibfnamefont {T.}~\bibnamefont
  {Harko}}, \bibinfo {author} {\bibfnamefont {Z.}~\bibnamefont {Kovacs}},\ and\
  \bibinfo {author} {\bibfnamefont {F.~S.~N.}\ \bibnamefont {Lobo}},\
  }\bibfield  {title} {\bibinfo {title} {{Thin accretion disk signatures in
  dynamical Chern-Simons modified gravity}},\ }\href
  {https://doi.org/10.1088/0264-9381/27/10/105010} {\bibfield  {journal}
  {\bibinfo  {journal} {Class. Quant. Grav.}\ }\textbf {\bibinfo {volume}
  {27}},\ \bibinfo {pages} {105010} (\bibinfo {year} {2010})},\ \Eprint
  {https://arxiv.org/abs/0909.1267} {arXiv:0909.1267 [gr-qc]} \BibitemShut
  {NoStop}%
\bibitem [{\citenamefont {Harko}\ \emph {et~al.}(2011)\citenamefont {Harko},
  \citenamefont {Kovacs},\ and\ \citenamefont {Lobo}}]{Harko:2010ua}%
  \BibitemOpen
  \bibfield  {author} {\bibinfo {author} {\bibfnamefont {T.}~\bibnamefont
  {Harko}}, \bibinfo {author} {\bibfnamefont {Z.}~\bibnamefont {Kovacs}},\ and\
  \bibinfo {author} {\bibfnamefont {F.~S.~N.}\ \bibnamefont {Lobo}},\
  }\bibfield  {title} {\bibinfo {title} {{Thin accretion disk signatures of
  slowly rotating black holes in Ho\v{r}ava gravity}},\ }\href
  {https://doi.org/10.1088/0264-9381/28/16/165001} {\bibfield  {journal}
  {\bibinfo  {journal} {Class. Quant. Grav.}\ }\textbf {\bibinfo {volume}
  {28}},\ \bibinfo {pages} {165001} (\bibinfo {year} {2011})},\ \Eprint
  {https://arxiv.org/abs/1009.1958} {arXiv:1009.1958 [gr-qc]} \BibitemShut
  {NoStop}%
\bibitem [{\citenamefont {Chakraborty}(2015)}]{Chakraborty:2014eha}%
  \BibitemOpen
  \bibfield  {author} {\bibinfo {author} {\bibfnamefont {S.}~\bibnamefont
  {Chakraborty}},\ }\bibfield  {title} {\bibinfo {title} {{Equilibrium
  configuration of perfect fluid orbiting around black holes in some classes of
  alternative gravity theories}},\ }\href
  {https://doi.org/10.1088/0264-9381/32/7/075007} {\bibfield  {journal}
  {\bibinfo  {journal} {Class. Quant. Grav.}\ }\textbf {\bibinfo {volume}
  {32}},\ \bibinfo {pages} {075007} (\bibinfo {year} {2015})},\ \Eprint
  {https://arxiv.org/abs/1406.0417} {arXiv:1406.0417 [gr-qc]} \BibitemShut
  {NoStop}%
\bibitem [{\citenamefont {Pun}\ \emph {et~al.}(2008{\natexlab{b}})\citenamefont
  {Pun}, \citenamefont {Kovacs},\ and\ \citenamefont {Harko}}]{Pun:2008ae}%
  \BibitemOpen
  \bibfield  {author} {\bibinfo {author} {\bibfnamefont {C.~S.~J.}\
  \bibnamefont {Pun}}, \bibinfo {author} {\bibfnamefont {Z.}~\bibnamefont
  {Kovacs}},\ and\ \bibinfo {author} {\bibfnamefont {T.}~\bibnamefont
  {Harko}},\ }\bibfield  {title} {\bibinfo {title} {{Thin accretion disks in
  f(R) modified gravity models}},\ }\href
  {https://doi.org/10.1103/PhysRevD.78.024043} {\bibfield  {journal} {\bibinfo
  {journal} {Phys. Rev. D}\ }\textbf {\bibinfo {volume} {78}},\ \bibinfo
  {pages} {024043} (\bibinfo {year} {2008}{\natexlab{b}})},\ \Eprint
  {https://arxiv.org/abs/0806.0679} {arXiv:0806.0679 [gr-qc]} \BibitemShut
  {NoStop}%
\bibitem [{\citenamefont {Li}\ \emph {et~al.}(2005)\citenamefont {Li},
  \citenamefont {Zimmerman}, \citenamefont {Narayan},\ and\ \citenamefont
  {McClintock}}]{Li:2004aq}%
  \BibitemOpen
  \bibfield  {author} {\bibinfo {author} {\bibfnamefont {L.-X.}\ \bibnamefont
  {Li}}, \bibinfo {author} {\bibfnamefont {E.~R.}\ \bibnamefont {Zimmerman}},
  \bibinfo {author} {\bibfnamefont {R.}~\bibnamefont {Narayan}},\ and\ \bibinfo
  {author} {\bibfnamefont {J.~E.}\ \bibnamefont {McClintock}},\ }\bibfield
  {title} {\bibinfo {title} {{Multi-temperature blackbody spectrum of a thin
  accretion disk around a Kerr black hole: Model computations and comparison
  with observations}},\ }\href {https://doi.org/10.1086/428089} {\bibfield
  {journal} {\bibinfo  {journal} {Astrophys. J. Suppl.}\ }\textbf {\bibinfo
  {volume} {157}},\ \bibinfo {pages} {335} (\bibinfo {year} {2005})},\ \Eprint
  {https://arxiv.org/abs/astro-ph/0411583} {arXiv:astro-ph/0411583}
  \BibitemShut {NoStop}%
\bibitem [{\citenamefont {Perlick}\ and\ \citenamefont
  {Tsupko}(2022)}]{Perlick:2021aok}%
  \BibitemOpen
  \bibfield  {author} {\bibinfo {author} {\bibfnamefont {V.}~\bibnamefont
  {Perlick}}\ and\ \bibinfo {author} {\bibfnamefont {O.~Y.}\ \bibnamefont
  {Tsupko}},\ }\bibfield  {title} {\bibinfo {title} {{Calculating black hole
  shadows: Review of analytical studies}},\ }\href
  {https://doi.org/10.1016/j.physrep.2021.10.004} {\bibfield  {journal}
  {\bibinfo  {journal} {Phys. Rept.}\ }\textbf {\bibinfo {volume} {947}},\
  \bibinfo {pages} {1} (\bibinfo {year} {2022})},\ \Eprint
  {https://arxiv.org/abs/2105.07101} {arXiv:2105.07101 [gr-qc]} \BibitemShut
  {NoStop}%
\bibitem [{\citenamefont {Psaltis}\ \emph
  {et~al.}(2020{\natexlab{a}})\citenamefont {Psaltis} \emph
  {et~al.}}]{EHT:2020qrl}%
  \BibitemOpen
  \bibfield  {author} {\bibinfo {author} {\bibfnamefont {D.}~\bibnamefont
  {Psaltis}} \emph {et~al.} (\bibinfo {collaboration} {Event Horizon
  Telescope}),\ }\bibfield  {title} {\bibinfo {title} {{Gravitational Test
  Beyond the First Post-Newtonian Order with the Shadow of the M87 Black
  Hole}},\ }\href {https://doi.org/10.1103/PhysRevLett.125.141104} {\bibfield
  {journal} {\bibinfo  {journal} {Phys. Rev. Lett.}\ }\textbf {\bibinfo
  {volume} {125}},\ \bibinfo {pages} {141104} (\bibinfo {year}
  {2020}{\natexlab{a}})},\ \Eprint {https://arxiv.org/abs/2010.01055}
  {arXiv:2010.01055 [gr-qc]} \BibitemShut {NoStop}%
\bibitem [{\citenamefont {Akiyama}\ \emph
  {et~al.}(2019{\natexlab{a}})\citenamefont {Akiyama} \emph
  {et~al.}}]{EHT-M87-1}%
  \BibitemOpen
  \bibfield  {author} {\bibinfo {author} {\bibfnamefont {K.}~\bibnamefont
  {Akiyama}} \emph {et~al.} (\bibinfo {collaboration} {Event Horizon
  Telescope}),\ }\bibfield  {title} {\bibinfo {title} {{First M87 Event Horizon
  Telescope Results. I. The Shadow of the Supermassive Black Hole}},\ }\href
  {https://doi.org/10.3847/2041-8213/ab0ec7} {\bibfield  {journal} {\bibinfo
  {journal} {Astrophys. J. Lett.}\ }\textbf {\bibinfo {volume} {875}},\
  \bibinfo {pages} {L1} (\bibinfo {year} {2019}{\natexlab{a}})},\ \Eprint
  {https://arxiv.org/abs/1906.11238} {arXiv:1906.11238 [astro-ph.GA]}
  \BibitemShut {NoStop}%
\bibitem [{\citenamefont {Akiyama}\ \emph
  {et~al.}(2019{\natexlab{b}})\citenamefont {Akiyama} \emph
  {et~al.}}]{EHT-M87-4}%
  \BibitemOpen
  \bibfield  {author} {\bibinfo {author} {\bibfnamefont {K.}~\bibnamefont
  {Akiyama}} \emph {et~al.} (\bibinfo {collaboration} {Event Horizon
  Telescope}),\ }\bibfield  {title} {\bibinfo {title} {{First M87 Event Horizon
  Telescope Results. IV. Imaging the Central Supermassive Black Hole}},\ }\href
  {https://doi.org/10.3847/2041-8213/ab0e85} {\bibfield  {journal} {\bibinfo
  {journal} {Astrophys. J. Lett.}\ }\textbf {\bibinfo {volume} {875}},\
  \bibinfo {pages} {L4} (\bibinfo {year} {2019}{\natexlab{b}})},\ \Eprint
  {https://arxiv.org/abs/1906.11241} {arXiv:1906.11241 [astro-ph.GA]}
  \BibitemShut {NoStop}%
\bibitem [{\citenamefont {Akiyama}\ \emph
  {et~al.}(2019{\natexlab{c}})\citenamefont {Akiyama}, \citenamefont {Alberdi},
  \citenamefont {Alef}, \citenamefont {Asada}, \citenamefont {Azulay},
  \citenamefont {Baczko}, \citenamefont {Ball}, \citenamefont {Balokovi{\'c}},
  \citenamefont {Barrett}, \citenamefont {Bintley} \emph {et~al.}}]{EHT-M87-5}%
  \BibitemOpen
  \bibfield  {author} {\bibinfo {author} {\bibfnamefont {K.}~\bibnamefont
  {Akiyama}}, \bibinfo {author} {\bibfnamefont {A.}~\bibnamefont {Alberdi}},
  \bibinfo {author} {\bibfnamefont {W.}~\bibnamefont {Alef}}, \bibinfo {author}
  {\bibfnamefont {K.}~\bibnamefont {Asada}}, \bibinfo {author} {\bibfnamefont
  {R.}~\bibnamefont {Azulay}}, \bibinfo {author} {\bibfnamefont {A.-K.}\
  \bibnamefont {Baczko}}, \bibinfo {author} {\bibfnamefont {D.}~\bibnamefont
  {Ball}}, \bibinfo {author} {\bibfnamefont {M.}~\bibnamefont {Balokovi{\'c}}},
  \bibinfo {author} {\bibfnamefont {J.}~\bibnamefont {Barrett}}, \bibinfo
  {author} {\bibfnamefont {D.}~\bibnamefont {Bintley}}, \emph {et~al.},\
  }\bibfield  {title} {\bibinfo {title} {First m87 event horizon telescope
  results. v. physical origin of the asymmetric ring},\ }\href@noop {}
  {\bibfield  {journal} {\bibinfo  {journal} {The Astrophysical Journal
  Letters}\ }\textbf {\bibinfo {volume} {875}},\ \bibinfo {pages} {L5}
  (\bibinfo {year} {2019}{\natexlab{c}})}\BibitemShut {NoStop}%
\bibitem [{\citenamefont {Akiyama}\ \emph
  {et~al.}(2019{\natexlab{d}})\citenamefont {Akiyama}, \citenamefont {Alberdi},
  \citenamefont {Alef}, \citenamefont {Asada}, \citenamefont {Azulay},
  \citenamefont {Baczko}, \citenamefont {Ball}, \citenamefont {Balokovi{\'c}},
  \citenamefont {Barrett}, \citenamefont {Bintley} \emph {et~al.}}]{EHT-M87-6}%
  \BibitemOpen
  \bibfield  {author} {\bibinfo {author} {\bibfnamefont {K.}~\bibnamefont
  {Akiyama}}, \bibinfo {author} {\bibfnamefont {A.}~\bibnamefont {Alberdi}},
  \bibinfo {author} {\bibfnamefont {W.}~\bibnamefont {Alef}}, \bibinfo {author}
  {\bibfnamefont {K.}~\bibnamefont {Asada}}, \bibinfo {author} {\bibfnamefont
  {R.}~\bibnamefont {Azulay}}, \bibinfo {author} {\bibfnamefont {A.-K.}\
  \bibnamefont {Baczko}}, \bibinfo {author} {\bibfnamefont {D.}~\bibnamefont
  {Ball}}, \bibinfo {author} {\bibfnamefont {M.}~\bibnamefont {Balokovi{\'c}}},
  \bibinfo {author} {\bibfnamefont {J.}~\bibnamefont {Barrett}}, \bibinfo
  {author} {\bibfnamefont {D.}~\bibnamefont {Bintley}}, \emph {et~al.},\
  }\bibfield  {title} {\bibinfo {title} {First m87 event horizon telescope
  results. vi. the shadow and mass of the central black hole},\ }\href@noop {}
  {\bibfield  {journal} {\bibinfo  {journal} {The Astrophysical Journal
  Letters}\ }\textbf {\bibinfo {volume} {875}},\ \bibinfo {pages} {L6}
  (\bibinfo {year} {2019}{\natexlab{d}})}\BibitemShut {NoStop}%
\bibitem [{\citenamefont {Akiyama}\ \emph
  {et~al.}(2022{\natexlab{a}})\citenamefont {Akiyama} \emph
  {et~al.}}]{EHT-SgrA-1}%
  \BibitemOpen
  \bibfield  {author} {\bibinfo {author} {\bibfnamefont {K.}~\bibnamefont
  {Akiyama}} \emph {et~al.} (\bibinfo {collaboration} {Event Horizon
  Telescope}),\ }\bibfield  {title} {\bibinfo {title} {{First Sagittarius A*
  Event Horizon Telescope Results. I. The Shadow of the Supermassive Black Hole
  in the Center of the Milky Way}},\ }\href
  {https://doi.org/10.3847/2041-8213/ac6674} {\bibfield  {journal} {\bibinfo
  {journal} {Astrophys. J. Lett.}\ }\textbf {\bibinfo {volume} {930}},\
  \bibinfo {pages} {L12} (\bibinfo {year} {2022}{\natexlab{a}})},\ \Eprint
  {https://arxiv.org/abs/2311.08680} {arXiv:2311.08680 [astro-ph.HE]}
  \BibitemShut {NoStop}%
\bibitem [{\citenamefont {Akiyama}\ \emph
  {et~al.}(2022{\natexlab{b}})\citenamefont {Akiyama}, \citenamefont {Alberdi},
  \citenamefont {Alef}, \citenamefont {Algaba}, \citenamefont {Anantua},
  \citenamefont {Asada}, \citenamefont {Azulay}, \citenamefont {Bach},
  \citenamefont {Baczko}, \citenamefont {Ball} \emph {et~al.}}]{EHT-SgrA-3}%
  \BibitemOpen
  \bibfield  {author} {\bibinfo {author} {\bibfnamefont {K.}~\bibnamefont
  {Akiyama}}, \bibinfo {author} {\bibfnamefont {A.}~\bibnamefont {Alberdi}},
  \bibinfo {author} {\bibfnamefont {W.}~\bibnamefont {Alef}}, \bibinfo {author}
  {\bibfnamefont {J.~C.}\ \bibnamefont {Algaba}}, \bibinfo {author}
  {\bibfnamefont {R.}~\bibnamefont {Anantua}}, \bibinfo {author} {\bibfnamefont
  {K.}~\bibnamefont {Asada}}, \bibinfo {author} {\bibfnamefont
  {R.}~\bibnamefont {Azulay}}, \bibinfo {author} {\bibfnamefont
  {U.}~\bibnamefont {Bach}}, \bibinfo {author} {\bibfnamefont {A.-K.}\
  \bibnamefont {Baczko}}, \bibinfo {author} {\bibfnamefont {D.}~\bibnamefont
  {Ball}}, \emph {et~al.},\ }\bibfield  {title} {\bibinfo {title} {First
  sagittarius a* event horizon telescope results. iii. imaging of the galactic
  center supermassive black hole},\ }\href@noop {} {\bibfield  {journal}
  {\bibinfo  {journal} {The Astrophysical Journal Letters}\ }\textbf {\bibinfo
  {volume} {930}},\ \bibinfo {pages} {L14} (\bibinfo {year}
  {2022}{\natexlab{b}})}\BibitemShut {NoStop}%
\bibitem [{\citenamefont {Akiyama}\ \emph
  {et~al.}(2022{\natexlab{c}})\citenamefont {Akiyama}, \citenamefont {Alberdi},
  \citenamefont {Alef}, \citenamefont {Algaba}, \citenamefont {Anantua},
  \citenamefont {Asada}, \citenamefont {Azulay}, \citenamefont {Bach},
  \citenamefont {Baczko}, \citenamefont {Ball} \emph {et~al.}}]{EHT-SgrA-4}%
  \BibitemOpen
  \bibfield  {author} {\bibinfo {author} {\bibfnamefont {K.}~\bibnamefont
  {Akiyama}}, \bibinfo {author} {\bibfnamefont {A.}~\bibnamefont {Alberdi}},
  \bibinfo {author} {\bibfnamefont {W.}~\bibnamefont {Alef}}, \bibinfo {author}
  {\bibfnamefont {J.~C.}\ \bibnamefont {Algaba}}, \bibinfo {author}
  {\bibfnamefont {R.}~\bibnamefont {Anantua}}, \bibinfo {author} {\bibfnamefont
  {K.}~\bibnamefont {Asada}}, \bibinfo {author} {\bibfnamefont
  {R.}~\bibnamefont {Azulay}}, \bibinfo {author} {\bibfnamefont
  {U.}~\bibnamefont {Bach}}, \bibinfo {author} {\bibfnamefont {A.-K.}\
  \bibnamefont {Baczko}}, \bibinfo {author} {\bibfnamefont {D.}~\bibnamefont
  {Ball}}, \emph {et~al.},\ }\bibfield  {title} {\bibinfo {title} {First
  sagittarius a* event horizon telescope results. iv. variability, morphology,
  and black hole mass},\ }\href@noop {} {\bibfield  {journal} {\bibinfo
  {journal} {The Astrophysical Journal Letters}\ }\textbf {\bibinfo {volume}
  {930}},\ \bibinfo {pages} {L15} (\bibinfo {year}
  {2022}{\natexlab{c}})}\BibitemShut {NoStop}%
\bibitem [{\citenamefont {Akiyama}\ \emph
  {et~al.}(2022{\natexlab{d}})\citenamefont {Akiyama}, \citenamefont {Alberdi},
  \citenamefont {Alef}, \citenamefont {Algaba}, \citenamefont {Anantua},
  \citenamefont {Asada}, \citenamefont {Azulay}, \citenamefont {Bach},
  \citenamefont {Baczko}, \citenamefont {Ball} \emph {et~al.}}]{EHT-SgrA-5}%
  \BibitemOpen
  \bibfield  {author} {\bibinfo {author} {\bibfnamefont {K.}~\bibnamefont
  {Akiyama}}, \bibinfo {author} {\bibfnamefont {A.}~\bibnamefont {Alberdi}},
  \bibinfo {author} {\bibfnamefont {W.}~\bibnamefont {Alef}}, \bibinfo {author}
  {\bibfnamefont {J.~C.}\ \bibnamefont {Algaba}}, \bibinfo {author}
  {\bibfnamefont {R.}~\bibnamefont {Anantua}}, \bibinfo {author} {\bibfnamefont
  {K.}~\bibnamefont {Asada}}, \bibinfo {author} {\bibfnamefont
  {R.}~\bibnamefont {Azulay}}, \bibinfo {author} {\bibfnamefont
  {U.}~\bibnamefont {Bach}}, \bibinfo {author} {\bibfnamefont {A.-K.}\
  \bibnamefont {Baczko}}, \bibinfo {author} {\bibfnamefont {D.}~\bibnamefont
  {Ball}}, \emph {et~al.},\ }\bibfield  {title} {\bibinfo {title} {First
  sagittarius a* event horizon telescope results. v. testing astrophysical
  models of the galactic center black hole},\ }\href@noop {} {\bibfield
  {journal} {\bibinfo  {journal} {The Astrophysical Journal Letters}\ }\textbf
  {\bibinfo {volume} {930}},\ \bibinfo {pages} {L16} (\bibinfo {year}
  {2022}{\natexlab{d}})}\BibitemShut {NoStop}%
\bibitem [{\citenamefont {Akiyama}\ \emph
  {et~al.}(2022{\natexlab{e}})\citenamefont {Akiyama}, \citenamefont {Alberdi},
  \citenamefont {Alef}, \citenamefont {Algaba}, \citenamefont {Anantua},
  \citenamefont {Asada}, \citenamefont {Azulay}, \citenamefont {Bach},
  \citenamefont {Baczko}, \citenamefont {Ball} \emph {et~al.}}]{EHT-SgrA-6}%
  \BibitemOpen
  \bibfield  {author} {\bibinfo {author} {\bibfnamefont {K.}~\bibnamefont
  {Akiyama}}, \bibinfo {author} {\bibfnamefont {A.}~\bibnamefont {Alberdi}},
  \bibinfo {author} {\bibfnamefont {W.}~\bibnamefont {Alef}}, \bibinfo {author}
  {\bibfnamefont {J.~C.}\ \bibnamefont {Algaba}}, \bibinfo {author}
  {\bibfnamefont {R.}~\bibnamefont {Anantua}}, \bibinfo {author} {\bibfnamefont
  {K.}~\bibnamefont {Asada}}, \bibinfo {author} {\bibfnamefont
  {R.}~\bibnamefont {Azulay}}, \bibinfo {author} {\bibfnamefont
  {U.}~\bibnamefont {Bach}}, \bibinfo {author} {\bibfnamefont {A.-K.}\
  \bibnamefont {Baczko}}, \bibinfo {author} {\bibfnamefont {D.}~\bibnamefont
  {Ball}}, \emph {et~al.},\ }\bibfield  {title} {\bibinfo {title} {First
  sagittarius a* event horizon telescope results. vi. testing the black hole
  metric},\ }\href@noop {} {\bibfield  {journal} {\bibinfo  {journal} {The
  Astrophysical Journal Letters}\ }\textbf {\bibinfo {volume} {930}},\ \bibinfo
  {pages} {L17} (\bibinfo {year} {2022}{\natexlab{e}})}\BibitemShut {NoStop}%
\bibitem [{\citenamefont {Afrin}\ \emph {et~al.}(2023)\citenamefont {Afrin},
  \citenamefont {Vagnozzi},\ and\ \citenamefont {Ghosh}}]{Afrin2023}%
  \BibitemOpen
  \bibfield  {author} {\bibinfo {author} {\bibfnamefont {M.}~\bibnamefont
  {Afrin}}, \bibinfo {author} {\bibfnamefont {S.}~\bibnamefont {Vagnozzi}},\
  and\ \bibinfo {author} {\bibfnamefont {S.~G.}\ \bibnamefont {Ghosh}},\
  }\bibfield  {title} {\bibinfo {title} {{Tests of Loop Quantum Gravity from
  the Event Horizon Telescope Results of Sgr A*}},\ }\href
  {https://doi.org/10.3847/1538-4357/acb334} {\bibfield  {journal} {\bibinfo
  {journal} {Astrophys. J.}\ }\textbf {\bibinfo {volume} {944}},\ \bibinfo
  {pages} {149} (\bibinfo {year} {2023})},\ \Eprint
  {https://arxiv.org/abs/2209.12584} {arXiv:2209.12584 [gr-qc]} \BibitemShut
  {NoStop}%
\bibitem [{\citenamefont {Vagnozzi}\ \emph {et~al.}(2023)\citenamefont
  {Vagnozzi} \emph {et~al.}}]{Vagnozzi2023}%
  \BibitemOpen
  \bibfield  {author} {\bibinfo {author} {\bibfnamefont {S.}~\bibnamefont
  {Vagnozzi}} \emph {et~al.},\ }\bibfield  {title} {\bibinfo {title}
  {{Horizon-scale tests of gravity theories and fundamental physics from the
  Event Horizon Telescope image of Sagittarius A}},\ }\href
  {https://doi.org/10.1088/1361-6382/acd97b} {\bibfield  {journal} {\bibinfo
  {journal} {Class. Quant. Grav.}\ }\textbf {\bibinfo {volume} {40}},\ \bibinfo
  {pages} {165007} (\bibinfo {year} {2023})},\ \Eprint
  {https://arxiv.org/abs/2205.07787} {arXiv:2205.07787 [gr-qc]} \BibitemShut
  {NoStop}%
\bibitem [{\citenamefont {Psaltis}\ \emph
  {et~al.}(2020{\natexlab{b}})\citenamefont {Psaltis} \emph
  {et~al.}}]{EHT2020qrl}%
  \BibitemOpen
  \bibfield  {author} {\bibinfo {author} {\bibfnamefont {D.}~\bibnamefont
  {Psaltis}} \emph {et~al.} (\bibinfo {collaboration} {Event Horizon
  Telescope}),\ }\bibfield  {title} {\bibinfo {title} {{Gravitational Test
  Beyond the First Post-Newtonian Order with the Shadow of the M87 Black
  Hole}},\ }\href {https://doi.org/10.1103/PhysRevLett.125.141104} {\bibfield
  {journal} {\bibinfo  {journal} {Phys. Rev. Lett.}\ }\textbf {\bibinfo
  {volume} {125}},\ \bibinfo {pages} {141104} (\bibinfo {year}
  {2020}{\natexlab{b}})},\ \Eprint {https://arxiv.org/abs/2010.01055}
  {arXiv:2010.01055 [gr-qc]} \BibitemShut {NoStop}%
\bibitem [{\citenamefont {Khodadi}\ \emph {et~al.}(2024)\citenamefont
  {Khodadi}, \citenamefont {Vagnozzi},\ and\ \citenamefont
  {Firouzjaee}}]{Khodadi:2024ubi}%
  \BibitemOpen
  \bibfield  {author} {\bibinfo {author} {\bibfnamefont {M.}~\bibnamefont
  {Khodadi}}, \bibinfo {author} {\bibfnamefont {S.}~\bibnamefont {Vagnozzi}},\
  and\ \bibinfo {author} {\bibfnamefont {J.~T.}\ \bibnamefont {Firouzjaee}},\
  }\bibfield  {title} {\bibinfo {title} {{Event Horizon Telescope observations
  exclude compact objects in baseline mimetic gravity}},\ }\href
  {https://doi.org/10.1038/s41598-024-78264-y} {\bibfield  {journal} {\bibinfo
  {journal} {Sci. Rep.}\ }\textbf {\bibinfo {volume} {14}},\ \bibinfo {pages}
  {26932} (\bibinfo {year} {2024})},\ \Eprint
  {https://arxiv.org/abs/2408.03241} {arXiv:2408.03241 [gr-qc]} \BibitemShut
  {NoStop}%
\bibitem [{\citenamefont {Collodel}\ \emph {et~al.}(2021)\citenamefont
  {Collodel}, \citenamefont {Doneva},\ and\ \citenamefont
  {Yazadjiev}}]{Collodel:2021gxu}%
  \BibitemOpen
  \bibfield  {author} {\bibinfo {author} {\bibfnamefont {L.~G.}\ \bibnamefont
  {Collodel}}, \bibinfo {author} {\bibfnamefont {D.~D.}\ \bibnamefont
  {Doneva}},\ and\ \bibinfo {author} {\bibfnamefont {S.~S.}\ \bibnamefont
  {Yazadjiev}},\ }\bibfield  {title} {\bibinfo {title} {{Circular Orbit
  Structure and Thin Accretion Disks around Kerr Black Holes with Scalar
  Hair}},\ }\href {https://doi.org/10.3847/1538-4357/abe305} {\bibfield
  {journal} {\bibinfo  {journal} {Astrophys. J.}\ }\textbf {\bibinfo {volume}
  {910}},\ \bibinfo {pages} {52} (\bibinfo {year} {2021})},\ \Eprint
  {https://arxiv.org/abs/2101.05073} {arXiv:2101.05073 [astro-ph.HE]}
  \BibitemShut {NoStop}%
\bibitem [{\citenamefont {Wu}\ \emph {et~al.}(2024)\citenamefont {Wu},
  \citenamefont {Feng},\ and\ \citenamefont {Chen}}]{Wu:2024sng}%
  \BibitemOpen
  \bibfield  {author} {\bibinfo {author} {\bibfnamefont {Y.}~\bibnamefont
  {Wu}}, \bibinfo {author} {\bibfnamefont {H.}~\bibnamefont {Feng}},\ and\
  \bibinfo {author} {\bibfnamefont {W.-Q.}\ \bibnamefont {Chen}},\ }\bibfield
  {title} {\bibinfo {title} {{Thin accretion disk around black hole in
  Einstein\textendash{}Maxwell-scalar theory}},\ }\href
  {https://doi.org/10.1140/epjc/s10052-024-13454-6} {\bibfield  {journal}
  {\bibinfo  {journal} {Eur. Phys. J. C}\ }\textbf {\bibinfo {volume} {84}},\
  \bibinfo {pages} {1075} (\bibinfo {year} {2024})},\ \Eprint
  {https://arxiv.org/abs/2410.14113} {arXiv:2410.14113 [gr-qc]} \BibitemShut
  {NoStop}%
\bibitem [{\citenamefont {Kurmanov}\ \emph {et~al.}(2024)\citenamefont
  {Kurmanov}, \citenamefont {Boshkayev}, \citenamefont {Konysbayev},
  \citenamefont {Luongo}, \citenamefont {Saiyp}, \citenamefont {Urazalina},
  \citenamefont {Ikhsan},\ and\ \citenamefont {Suliyeva}}]{Kurmanov:2024hpn}%
  \BibitemOpen
  \bibfield  {author} {\bibinfo {author} {\bibfnamefont {Y.}~\bibnamefont
  {Kurmanov}}, \bibinfo {author} {\bibfnamefont {K.}~\bibnamefont {Boshkayev}},
  \bibinfo {author} {\bibfnamefont {T.}~\bibnamefont {Konysbayev}}, \bibinfo
  {author} {\bibfnamefont {O.}~\bibnamefont {Luongo}}, \bibinfo {author}
  {\bibfnamefont {N.}~\bibnamefont {Saiyp}}, \bibinfo {author} {\bibfnamefont
  {A.}~\bibnamefont {Urazalina}}, \bibinfo {author} {\bibfnamefont
  {G.}~\bibnamefont {Ikhsan}},\ and\ \bibinfo {author} {\bibfnamefont
  {G.}~\bibnamefont {Suliyeva}},\ }\bibfield  {title} {\bibinfo {title}
  {{Accretion disks properties around regular black hole solutions obtained
  from non-linear electrodynamics}},\ }\href
  {https://doi.org/10.1016/j.dark.2024.101566} {\bibfield  {journal} {\bibinfo
  {journal} {Phys. Dark Univ.}\ }\textbf {\bibinfo {volume} {46}},\ \bibinfo
  {pages} {101566} (\bibinfo {year} {2024})},\ \Eprint
  {https://arxiv.org/abs/2404.15437} {arXiv:2404.15437 [gr-qc]} \BibitemShut
  {NoStop}%
\bibitem [{\citenamefont {Liu}\ \emph {et~al.}(2024{\natexlab{c}})\citenamefont
  {Liu}, \citenamefont {He}, \citenamefont {Liu}, \citenamefont {Han},\ and\
  \citenamefont {Yang}}]{Liu:2024brf}%
  \BibitemOpen
  \bibfield  {author} {\bibinfo {author} {\bibfnamefont {A.}~\bibnamefont
  {Liu}}, \bibinfo {author} {\bibfnamefont {T.-Y.}\ \bibnamefont {He}},
  \bibinfo {author} {\bibfnamefont {M.}~\bibnamefont {Liu}}, \bibinfo {author}
  {\bibfnamefont {Z.-W.}\ \bibnamefont {Han}},\ and\ \bibinfo {author}
  {\bibfnamefont {R.-J.}\ \bibnamefont {Yang}},\ }\bibfield  {title} {\bibinfo
  {title} {{Possible signatures of higher dimension in thin accretion disk
  around brane world black hole}},\ }\href
  {https://doi.org/10.1088/1475-7516/2024/07/062} {\bibfield  {journal}
  {\bibinfo  {journal} {JCAP}\ }\textbf {\bibinfo {volume} {07}},\ \bibinfo
  {pages} {062}},\ \Eprint {https://arxiv.org/abs/2404.14131} {arXiv:2404.14131
  [gr-qc]} \BibitemShut {NoStop}%
\bibitem [{\citenamefont {Lee}\ \emph {et~al.}(2023)\citenamefont {Lee},
  \citenamefont {Hu}, \citenamefont {Guo},\ and\ \citenamefont
  {Chen}}]{Lee:2022rtg}%
  \BibitemOpen
  \bibfield  {author} {\bibinfo {author} {\bibfnamefont {T.-C.}\ \bibnamefont
  {Lee}}, \bibinfo {author} {\bibfnamefont {Z.}~\bibnamefont {Hu}}, \bibinfo
  {author} {\bibfnamefont {M.}~\bibnamefont {Guo}},\ and\ \bibinfo {author}
  {\bibfnamefont {B.}~\bibnamefont {Chen}},\ }\bibfield  {title} {\bibinfo
  {title} {{Circular orbits and polarized images of charged particles orbiting
  a Kerr black hole with a weak magnetic field}},\ }\href
  {https://doi.org/10.1103/PhysRevD.108.024008} {\bibfield  {journal} {\bibinfo
   {journal} {Phys. Rev. D}\ }\textbf {\bibinfo {volume} {108}},\ \bibinfo
  {pages} {024008} (\bibinfo {year} {2023})},\ \Eprint
  {https://arxiv.org/abs/2211.04143} {arXiv:2211.04143 [gr-qc]} \BibitemShut
  {NoStop}%
\bibitem [{\citenamefont {Asuk\"ula}\ \emph {et~al.}(2024)\citenamefont
  {Asuk\"ula}, \citenamefont {Hohmann}, \citenamefont {Karanasou},
  \citenamefont {Bahamonde}, \citenamefont {Pfeifer},\ and\ \citenamefont
  {Rosa}}]{Asukula:2023akj}%
  \BibitemOpen
  \bibfield  {author} {\bibinfo {author} {\bibfnamefont {H.}~\bibnamefont
  {Asuk\"ula}}, \bibinfo {author} {\bibfnamefont {M.}~\bibnamefont {Hohmann}},
  \bibinfo {author} {\bibfnamefont {V.}~\bibnamefont {Karanasou}}, \bibinfo
  {author} {\bibfnamefont {S.}~\bibnamefont {Bahamonde}}, \bibinfo {author}
  {\bibfnamefont {C.}~\bibnamefont {Pfeifer}},\ and\ \bibinfo {author}
  {\bibfnamefont {J.~a.~L.}\ \bibnamefont {Rosa}},\ }\bibfield  {title}
  {\bibinfo {title} {{Spherically symmetric vacuum solutions in one-parameter
  new general relativity and their phenomenology}},\ }\href
  {https://doi.org/10.1103/PhysRevD.109.064027} {\bibfield  {journal} {\bibinfo
   {journal} {Phys. Rev. D}\ }\textbf {\bibinfo {volume} {109}},\ \bibinfo
  {pages} {064027} (\bibinfo {year} {2024})},\ \Eprint
  {https://arxiv.org/abs/2311.17999} {arXiv:2311.17999 [gr-qc]} \BibitemShut
  {NoStop}%
\bibitem [{\citenamefont {Olmo}\ \emph {et~al.}(2023)\citenamefont {Olmo},
  \citenamefont {Rosa}, \citenamefont {Rubiera-Garcia},\ and\ \citenamefont
  {Saez-Chillon~Gomez}}]{Olmo:2023lil}%
  \BibitemOpen
  \bibfield  {author} {\bibinfo {author} {\bibfnamefont {G.~J.}\ \bibnamefont
  {Olmo}}, \bibinfo {author} {\bibfnamefont {J.~L.}\ \bibnamefont {Rosa}},
  \bibinfo {author} {\bibfnamefont {D.}~\bibnamefont {Rubiera-Garcia}},\ and\
  \bibinfo {author} {\bibfnamefont {D.}~\bibnamefont {Saez-Chillon~Gomez}},\
  }\bibfield  {title} {\bibinfo {title} {{Shadows and photon rings of regular
  black holes and geonic horizonless compact objects}},\ }\href
  {https://doi.org/10.1088/1361-6382/aceacd} {\bibfield  {journal} {\bibinfo
  {journal} {Class. Quant. Grav.}\ }\textbf {\bibinfo {volume} {40}},\ \bibinfo
  {pages} {174002} (\bibinfo {year} {2023})},\ \Eprint
  {https://arxiv.org/abs/2302.12064} {arXiv:2302.12064 [gr-qc]} \BibitemShut
  {NoStop}%
\bibitem [{\citenamefont {Luminet}(1979)}]{Luminet:1979nyg}%
  \BibitemOpen
  \bibfield  {author} {\bibinfo {author} {\bibfnamefont {J.~P.}\ \bibnamefont
  {Luminet}},\ }\bibfield  {title} {\bibinfo {title} {{Image of a spherical
  black hole with thin accretion disk}},\ }\href@noop {} {\bibfield  {journal}
  {\bibinfo  {journal} {Astron. Astrophys.}\ }\textbf {\bibinfo {volume}
  {75}},\ \bibinfo {pages} {228} (\bibinfo {year} {1979})}\BibitemShut
  {NoStop}%
\bibitem [{\citenamefont {Huang}\ \emph {et~al.}(2023)\citenamefont {Huang},
  \citenamefont {Guo}, \citenamefont {Cui}, \citenamefont {Jiang},\ and\
  \citenamefont {Lin}}]{Huang:2023ilm}%
  \BibitemOpen
  \bibfield  {author} {\bibinfo {author} {\bibfnamefont {Y.-X.}\ \bibnamefont
  {Huang}}, \bibinfo {author} {\bibfnamefont {S.}~\bibnamefont {Guo}}, \bibinfo
  {author} {\bibfnamefont {Y.-H.}\ \bibnamefont {Cui}}, \bibinfo {author}
  {\bibfnamefont {Q.-Q.}\ \bibnamefont {Jiang}},\ and\ \bibinfo {author}
  {\bibfnamefont {K.}~\bibnamefont {Lin}},\ }\bibfield  {title} {\bibinfo
  {title} {{Influence of accretion disk on the optical appearance of the
  Kazakov-Solodukhin black hole}},\ }\href
  {https://doi.org/10.1103/PhysRevD.107.123009} {\bibfield  {journal} {\bibinfo
   {journal} {Phys. Rev. D}\ }\textbf {\bibinfo {volume} {107}},\ \bibinfo
  {pages} {123009} (\bibinfo {year} {2023})},\ \Eprint
  {https://arxiv.org/abs/2311.00302} {arXiv:2311.00302 [gr-qc]} \BibitemShut
  {NoStop}%
\end{thebibliography}%
\end{document}